\documentclass[pra,aps,twocolumn,floatfix]{revtex4-1}
\usepackage{graphicx,graphics,psfrag,amsmath,calc}
\usepackage{epsfig}
\usepackage{color, comment}
\usepackage{mathbbol}
\usepackage{float}
\usepackage{placeins}
\usepackage{subfigure}

\begin{document}

\title{Eigenspectrum, Chern Numbers and Phase Diagrams \\
  of Ultracold Color-orbit Coupled ${\rm SU(3)}$ Fermions
  in Optical Lattices}

\author{Man Hon Yau and C. A. R. S\'a de Melo}

\affiliation{
School of Physics, Georgia Institute of Technology, 
Atlanta, 30332, USA
}

\date{\today}

\begin{abstract}
We study ultracold color fermions with three internal states Red, Green and Blue with
${\rm SU(3)}$ symmetry in optical lattices, when color-orbit coupling and color-flip fields
are present.  This system corresponds to a generalization of two-internal state fermions
with ${\rm SU(2)}$ symmetry in the presence of spin-orbit coupling and spin-flipping Zeeman
fields. We investigate the eigenspectrum and Chern numbers to describe different topological
phases that emerge in the phase diagrams of color-orbit coupled fermions in optical lattices. 
We obtain the phases as a function of artificial magnetic, color-orbit and color-flip fields
that can be independently controlled.  For fixed artificial magnetic flux ratio, we identify
topological quantum phases and phase transitions in the phase diagrams of chemical
potential versus color-flip fields or color-orbit coupling, where the chirality and number of
midgap edge states changes.
The topologically non-trivial phases are classified in three groups:
the first group has total non-zero chirality and exhibit only the quantum charge Hall effect;
the second group has total non-zero chirality and exhibit both quantum charge and quantum
color Hall effects;
and the third group has total zero chirality, but exhibit the quantum color Hall effect.
These phases are generalizations of the
quantum Hall and quantum spin Hall phases for charged spin-$1/2$ fermions.
Lastly, we also describe the color density of states and a staircase structure in the total
and color filling factors versus chemical potential for fixed color-orbit,
color-flip and magnetic flux ratio. We show the existence of incompressible states 
at rational filling factors precisely given by a gap-labelling theorem that relates the filling
factors to the magnetic flux ratio and topological quantum numbers.
\end{abstract}

\maketitle

%
%

%
\section{Introduction}
\label{sec:introduction}

Ultracold fermions loaded in optical lattices have become ideal systems to 
study related electronic phase diagrams and transport properties, because they provide
a clean and well controlled playground to change various lattice parameters and external fields
at the turn of a knob. While several experimental groups have worked mostly with Fermi isotopes
$^6$Li and $^{40}$K using two internal states to study various aspects of interacting ${\rm SU(2)}$
fermions, there has been a growing interest in studying
${\rm SU(N)}$ generalizations of these systems.
Examples of atomic ${\rm SU(N)}$  fermions found in nature are fermionic isotopes of
closed shell atoms with two electrons in their outer electronic configuration.
Two systems have been studied by several groups,
one of them is $^{173}{\rm Yb}$, a fermionic isotope of Ytterbium and the other
is $^{87} {\rm Sr}$, a fermionic isotope of Strontium.

The fermionic isotope $^{173}$Yb has electronic shell
structure $[{\rm Xe}] 4{\rm f}^{14}6{\rm s}^2$, with electronic spin ${\rm S} = 0$ and nu-
clear spin $ {\rm I} = 5/2$. The electronic ground state of $^{173} {\rm Yb}$
is $^1{\rm S}_0$, which is six-fold degenerate because of its nuclear
spin. The six degenerate states have nuclear spin projec-
tions $ m_{\rm I} = \{ \pm 5/2, \pm 3/2, \pm 1/2 \}$. Atoms in any selected
state can be manipulated out of a trap or transformed into a desired
nuclear spin state, so that the ground state of trapped $^{173} {\rm Yb}$ can be up to
six-fold degenerate~\cite{takahashi-2007, ueda-2009, takahashi-2010,takahashi-2012, fallani-2014, folling-2016}.

The fermionic isotope $^{87} {\rm Sr}$ has electronic shell struc-
ture $ [{\rm Kr}] 5{\rm s}^2$, with electronic spin ${\rm S}= 0$ and nuclear
spin ${\rm I} = 9/2$. The electronic ground state of $^{87}{\rm Sr}$ is $^1{\rm S}_0$,
which is ten-fold degenerate because of its nuclear spin.
The ten degenerate states have nuclear spin projection
$m_{\rm I} = \{ \pm 9/2, \pm 7/2, \pm 5/2, \pm 3/2, \pm1/2 \}$. Again, atoms in
any selected state can be manipulated out of a trap or
transformed into a desired nuclear spin state, so that the
ground state of trapped $^{87}{\rm Sr}$ can be up to
ten-fold degenerate~\cite{killian-2010, schreck-2010, schreck-2011, killian-2013, killian-2014}.
In addition, interactions between these close shell
atoms are independent of their nuclear spin states at the
atomic energy scales of interest, and therefore interac-
tions are SU(N) symmetric. Since experiments are conducted
at very low temperatures, the collisional
properties of these atoms are dominated by s-wave scattering,
and the interactions are local in space, that is, they
are contact interactions described by a delta function potential that
is independent of the nuclear spin states of the atoms.
As a result, $^{173}{\rm Yb}$ can be up to ${\rm SU(6)}$ symmetric,
while $^{87}{\rm Sr}$ can be up to ${\rm SU(10)}$ symmetric in their nuclear spin projections.
In addition, orbital-Feshbach resonances can be used to control the strength of
the SU(N)-symmetric interactions from weak to strong~\cite{fallani-2015, folling-2015}.
Since any three nuclear states of $^{173}{\rm Yb}$ or $^{87}{\rm Sr}$ can
be selected and trapped in an optical lattice, we label these nuclear states by color
$\{R,G,B\}$ or pseudo-spin $\{\uparrow, 0, \downarrow\}$ to describe a Fermi system with
${\rm SU(3)}$ symmetry.

It is now possible to create artificial magnetic fields~\cite{spielman-2009}
in optical lattices~\cite{bloch-2013, ketterle-2013} that mimic electronic materials exhibiting  
integer~\cite{klitzing-1980} and fractional~\cite{stormer-1982} 
quantum Hall effects. The synthetic magnetic flux values created are 
sufficiently large to allow for the experimental exploration 
of the intricacies of the Harper's model~\cite{harper-1955} and the Hofstadter 
butterfly~\cite{hofstadter-1976}, as well as the experimental determination
of Chern numbers~\cite{bloch-2015}. In addition, artificial magnetic fields
for ${\rm SU(2)}$ fermions in optical lattices could be used to simulate the phenomenon
of magnetic field induced reentrant superfluidity, as discussed in the context of superconductivity
in condensed matter physics for spin-$1/2$ fermions in standard
lattices~\cite{sademelo-1993, sademelo-1994, sademelo-1996, sademelo-1998}.
Furthermore, the creation of artificial spin-orbit coupling
for ultra-cold atoms~\cite{spielman-2011} also allows for  
the simulation of electronic materials exhibiting the quantum spin-Hall 
effect~\cite{kane-2005,haldane-2005,zhang-2006}.  
  
For ultracold fermions in optical lattices, artificial magnetic fields enable 
studies of topological insulators that break time reversal symmetry, 
such as quantum hall systems, while artificial spin-orbit fields allow  
for studies of topological insulators that do not break time reversal symmetry, 
such as quantum spin-Hall systems. Both types of topological insulators are
characterized by Berry curvatures and Chern numbers, which have been 
measured experimentally using time of flight techniques~\cite{weitenberg-2016}, 
inspired by theoretical proposals~\cite{indu-2011, lewenstein-2014}, and
using dynamics of the center of mass of the atomic cloud~\cite{esslinger-2014},
also motivated by theoretical work~\cite{cooper-2012, goldman-2013a}.
However, studies of ultracold fermions may go beyond the quantum simulation
of spin-$1/2$ topological insulators under typical condensed matter
conditions~\cite{goldman-2012}, because artificial magnetic, spin-orbit and Zeeman
fields may be adjusted independently~\cite{yau-2019}.

Artificial magnetic, spin-orbit and Zeeman fields in spin-$1/2$ ultracold Fermi atoms may be
independently tunned via a combination 
of experimental techniques that produce artificial magnetic fluxes without 
using internal states, such as laser assisted 
tunneling~\cite{bloch-2013, ketterle-2013},
and that produce spin-orbit and Zeeman fields using internal states, such 
as Raman processes~\cite{spielman-2011} or 
radio-frequency chips~\cite{spielman-2010, roady-2020}.
These techniques can also be applied to ${\rm SU(3)}$ fermions with three internal states (colors)
and allow for the investigation of exotic topological insulating phases that
arise in optical lattices when artificial magnetic, color-orbit and color-flip fields are varied.
The present system in optical lattices expands the realm of phases
beyond Fermi liquid and superfluid for ${\rm SU(3)}$ fermions in the presence
of color-orbit and color-flip fields analyzed in the continuum
or in harmonic traps~\cite{kurkcuoglu-2018a, kurkcuoglu-2018b}. 

In this manuscript, we study the interplay of artificial magnetic, color-orbit and color-flip
fields for ultracold ${\rm SU(3)}$ fermions with three internal states (colors) and their effects on
topological insulators in regimes that cannot be reached or found in condensed matter physics.
We investigate the eigenspectrum and Chern numbers to describe different topological phases
that emerge in the phase diagrams of color-orbit coupled fermions in optical lattices. 
We obtain the phases as a function of artificial magnetic, color-orbit and color-flip fields
that can be independently controlled.  For a fixed artificial magnetic flux ratio, we identify
the topological quantum phases and phase transitions in the phase diagrams of
chemical potential versus color-flip fields or color-orbit coupling, where the chirality
and number of midgap edge states change.
The topologically non-trivial phases are classified in three groups:
the first group has total non-zero chirality and exhibit only the quantum charge Hall effect;
the second group has total non-zero chirality and exhibit both quantum charge and quantum
color Hall effects;
and the third group has total zero chirality, but exhibit the quantum color Hall effect.
These phases are generalizations of the
quantum Hall and quantum spin Hall phases for charged spin-$1/2$ fermions.
Lastly, we also describe the color density of states and 
a staircase structure in the total and color filling factors versus chemical potential
for fixed color-orbit, color-flip and magnetic flux ratio. We show the existence of
incompressible states at rational filling factors precisely given by a gap-labelling theorem
that relates the filling factors to the magnetic flux ratio and topological numbers.

The remainder of this manuscript is organized as follows.
In Sec.~\ref{sec:hamiltonian}, we describe the three-color Hamiltonian for ultracold fermions
loaded into a square optical lattice and in the presence of artificial magnetic, color-orbit
and color-flip fields.
In Sec.~\ref{sec:eigenspectrum}, we analyze the energy spectrum obtained from a generalized
Harper's matrix with open boundary conditions and obtain the energy dispersions for
bulk and edge states.
In Sec.~\ref{sec:color-chern-numbers}, we discuss the color Chern numbers to classify 
the insulating phases in the {\it charge} sector. We compute the Chern numbers via the
Berry curvatures associated with the eigenstates of the Hamiltonian with periodic
boundary conditions. We confirm the existence of a bulk-edge correspondence by
comparing the Chern number calculated via periodic boundary conditions
to the number of chiral edge states obtained via open boundary conditions.
In Sec.~\ref{sec:phase-diagrams},
we describe the phase diagrams of chemical potential
versus color-flip field (Sec.~\ref{sec:chemical-potential-vs-color-flip-field}) ,
the gap labelling theorem (Sec.~\ref{sec:gap-labelling-theorem}),
and the phase diagrams of chemical potential versus color-orbit coupling
(Sec.~\ref{sec:chemical-potential-vs-color-orbit-coupling}).
We identify phases that exhibit quantum charge Hall and quantum color Hall effects
in analogy to the quantum Hall effect and quantum spin Hall effect for spin-$1/2$ systems,
and we find phases that exhibit simultaneously quantum charge and color Hall effect,
which do not exist in spin-$1/2$ systems.
In Sec.~\ref{sec:color-density-of-states}, we analyze the color density of states for
the system with periodic boundary conditions, and show that the bulk gaps match
precisely with the gaps obtained for open boundary conditions. Furthermore, we compute
directly the filling factor as a function of the chemical potential and reveal the existence
of filling factor steps at the precise values given by the
gap-labelling theorem discussed in Sec.~\ref{sec:gap-labelling-theorem}
to describe insulating states.
Finally, in Sec.~\ref{sec:summary-and-conclusions}, we summarize our results
and state our conclusions.

\section{Three-Color Hamiltonian}
\label{sec:hamiltonian}

To discuss the phase diagrams and Chern numbers of colored fermions with three
internal states Red $(R)$, Green $(G)$ and Blue $(B)$, we consider these fermions
to be trapped in a two-dimensional square optical lattice. The Hamiltonian matrix 
or ultra-cold atoms with three internal states is 
\begin{eqnarray}
 \label{eqn:hamiltonian-matrix}
{\hat {\bf H}}
= 
\left(
\begin{array}{c c c }
\varepsilon_{R} ( \hat{\bf k}) & -h_x/\sqrt{2}   &0  \\
-h_x/\sqrt{2}   &  \varepsilon_{G} (\hat {\bf k}) & -h_x/\sqrt{2} \\
 0              &  -h_x/\sqrt{2} &    \varepsilon_{B} (\hat {\bf k})
\end{array}
\right),
\end{eqnarray}
when written in first quantization.
In Eq.~(\ref{eqn:hamiltonian-matrix}),  the term
\begin{equation}
\label{eqn:red-kinetic-energy}
\varepsilon_{R} (\hat{\bf k}) =
-2t \{\cos[({\hat k}_x - k_{T})a] + \cos[({\hat k}_y  - {\cal A}_y)a]\}
\end{equation}
corresponds to the kinetic energy of the $R$ state including the momentum transfer
$+k_T$ along the $x$ direction, arising from counter propagating Raman
beams~\cite{spielman-2011} or radio-frequency chips~\cite{spielman-2010, roady-2020},
and the vector potential ${\cal A}_y$ along the $y$ direction,  arising from
laser assistant tunneling~\cite{bloch-2013, ketterle-2013}.
The term 
\begin{equation}
\label{eqn:green-kinetic-energy}
\varepsilon_{G} (\hat{\bf k}) =
-2t \{\cos ({\hat k}_x) + \cos[({\hat k}_y  -{\cal A}_y)a]\}
\end{equation}
corresponds to the kinetic energy of the $G$ state, which experiences no momentum transfer,
but feels the presence of ${\cal A}_y$, 
and
\begin{equation}
\label{eqn:blue-kinetic-energy}
\varepsilon_{B} (\hat{\bf k}) =
-2t \{\cos[({\hat k}_x + k_{T})a] + \cos[({\hat k}_y  -{\cal A}_y)a]\}
\end{equation}
corresponds to the kinetic energy of $B$ state, including the momentum transfer
$-k_T$ along the $x$ direction, and the vector potential ${\cal A}_y$ along the $y$ direction.

In Eqs.~(\ref{eqn:red-kinetic-energy}), (\ref{eqn:green-kinetic-energy}),
and (\ref{eqn:blue-kinetic-energy}),  the parameter $t$ is the hopping amplitude, 
$a$ is the lattice spacing,  $k_T$ is the color-dependent momentum transfer
along the $x$ direction (artificial unidirectional 
color-orbit coupling), and ${\cal A}_y = eHx/\hbar c$ plays the role 
of the $y$-component of the artificial vector potential, 
where $H$ is identified as a synthetic magnetic field along the $z$-axis. 
Notice that ${\cal A}_y$ has dimensions of inverse length.
It is important to emphasize that the system is neutral, so there is no
charge $e$, that is, ${\cal A}_y a$ should be just viewed as a position
dependent phase $\phi (x) = {\cal A}_y a$. 
Lastly, $h_x$ represents a color-flip field along the $x$-direction, 
whose physical origin is a Rabi term that couples the Red and Green,
as well as the Green and Blue internal states of the atom. 
As described in Sec.~\ref{sec:introduction}, the vector potential ${\cal A}_y$ may be generated by 
laser assisted tunneling~\cite{bloch-2013, ketterle-2013}, while the 
color-dependent momentum transfer $k_T$ and color-flip field $h_x$ may be 
created via counter-propagating Raman beams~\cite{spielman-2011} or via 
radio-frequency chips~\cite{spielman-2010, roady-2020}.

The Hamiltonian matrix in Eq.~(\ref{eqn:hamiltonian-matrix}) acts on
a three-color wavefunction 
$
{\boldsymbol \Psi}({\bf r}) 
= 
\left[ \Psi_{R} ({\bf r}), \Psi_{G}({\bf r}), \Psi_{B} ({\bf r})  \right]^T,
$
where $T$ indicates transposition and ${\bf r} = (x, y)$ labels the coordinates
in the square lattice. An analogy to pseudo-spin-1 fermions or spin-1 bosons in 
optical lattices can be made by 
rewriting Eq.~(\ref{eqn:hamiltonian-matrix}) in terms of spin-1 matrices
${\bf J}_{\ell}$, with $\ell = \{x, y, z \}$ as
\begin{equation}
\label{eqn:hamiltonian-matrix-spin-1}  
{\hat {\bf H}} 
= 
\varepsilon_{G} ( \hat {\bf k} ) {\bf 1}
- h_x {\bf J}_x 
- h_z (\hat {\bf k}) {\bf J}_z 
+ g_z (\hat {\bf k} ){\bf J}_z^2 
\end{equation}
where 
$h_x$ plays the role of a Zeeman field along the $x$ axis in spin-space,
$
h_z ({\hat {\bf k}}) 
= 
\left[ 
\varepsilon_{B} ({\hat {\bf k}})
-
\varepsilon_{R} ({\hat {\bf k}})
\right]/2
$
represents momentum dependent Zeeman field along the
$z$ axis in spin-space, and 
$
g_z 
= 
\left[
\varepsilon_{B} ({\hat {\bf k}})
+
\varepsilon_{R} ({\hat {\bf k}})
\right]/2
- 
\varepsilon_{G} ({\hat {\bf k}})
$
describes a momentum dependent quadratic Zeeman shift
along the $z$ axis in spin-space, and thus can be viewed as a spin (color) quadrupolar effect.
The color states $\{R, G, B\}$ 
are directly mapped into pseudo-spin-1 states $\{ \uparrow, 0, \downarrow \}$. 
Notice that the presence of the color fields $h_x$, 
$h_z (\hat {\bf k})$ and $g_z (\hat {\bf k})$ breaks the SU(3) symmetry
of otherwise degenerate color bands.
To make some connections to quantum chromodynamics (QCD), we note
that the independent-particle Hamiltonian described in
Eqs.~(\ref{eqn:hamiltonian-matrix}) or~(\ref{eqn:hamiltonian-matrix-spin-1})  
in general does not commute with the Gell-Mann matrices ${\boldsymbol \lambda}_j$, 
which are the eight generators of ${\rm SU(3)}$. To visualize this clearly,
it is sufficient to recall that the angular momentum matrices ${\bf J}_\ell$
can be written in terms of ${\boldsymbol \lambda}_ j$ as 
$
{\bf J}_x 
= 
\left(
{\boldsymbol \lambda}_1 + {\boldsymbol \lambda}_6
\right)/2;
$
$
{\bf J}_y 
= 
\left( 
{\boldsymbol \lambda}_2 + {\boldsymbol \lambda}_7
\right)/2;
$
and
$
{\bf J}_z 
= 
\left(
{\boldsymbol \lambda}_3 + \sqrt{3}{\boldsymbol \lambda}_8
\right)/2
$
and to show that the commutator 
$
\left[ 
{\hat {\bf H} }, {\boldsymbol \lambda}_j 
\right]
\ne 0
$.
The Hamiltonian in
Eqs.~(\ref{eqn:hamiltonian-matrix}) or~(\ref{eqn:hamiltonian-matrix-spin-1})  
becomes ${\rm SU(3)}$ invariant only when the fields
$h_x  = h_z ({\hat {\bf k}}) = g_z ({\hat {\bf k}}) = 0$, 
rendering ${\hat {\bf H}}$ diagonal and proportional to the unit
matrix {\bf 1}, that is, all color states become degenerate. 

Having described the Hamiltonian of our system in this section, we discuss next
the eigenspectrum associated with the Hamiltonian matrix described in
Eqs.~(\ref{eqn:hamiltonian-matrix}) or~(\ref{eqn:hamiltonian-matrix-spin-1}).  

\section{Harper's Eigenspectrum}
\label{sec:eigenspectrum}

To obtain the eigenspectrum, it is important to establish the boundary conditions.
We work in a cylindrical geometry having finite number $N$ of sites along the
$x$-direction but periodic  boundary conditions along the $y$-direction. In the present case, 
$k_y$ is a good quantum number, while $k_x$ is not, and 
color-dependent Harper's matrix 
\begin{eqnarray}
\label{eqn:hamiltonian-matrix-cylinder-geometry}
{\bf H} 
=
\left(
\begin{array}{c c c c c}
{\bf A}_{m-2}  & {\bf B}       &  {\bf 0}      &  {\bf 0}      & {\bf 0}       \\
{\bf B}^*      & {\bf A}_{m-1} &  {\bf B}      &  {\bf 0}      & {\bf 0}       \\
{\bf 0}        & {\bf B}^*     &  {\bf A}_m    & {\bf B}       & {\bf 0}       \\
{\bf 0}        & {\bf 0}       &  {\bf B}^*    & {\bf A}_{m+1} & {\bf B}       \\ 
{\bf 0}        & {\bf 0}       &  {\bf 0}      & {\bf B}^*     & {\bf A}_{m+2} \\
\end{array}
\right)
\end{eqnarray}
has a tridiagonal block structure that couples neighboring sites 
$(m-1, m, m+1)$ along the $x$-direction, but possesses discrete 
translational invariance along the $y$-axis. This is a generalization of the
Harper's matrix for spin-$1/2$ fermions with two internal states~\cite{harper-1955}.
The matrices ${\bf A}$, ${\bf B}$ and the null matrix ${\bf 0}$ 
consist of $3\times3$ blocks with entries labeled by internal 
color states $\{R, G, B\}$ or pseudo-spin-1 states 
$(\{\uparrow, 0, \downarrow \})$. 
The size of the space labeled by the site index $m$ is $N$, thus 
the total dimension of the matrix ${\bf H}$ in 
Eq.~(\ref{eqn:hamiltonian-matrix-cylinder-geometry})
is $3N \times 3N$. The matrix indexed by position $x = m a$ is   
\begin{eqnarray}
{\bf A}_m 
=
\left(
\begin{array}{ccc}
{\bf A}_{m R} & -h_x/\sqrt{2}   & 0\\
-h_x/\sqrt{2}        & {\bf A}_{m G}    & -h_x/\sqrt{2}\\
0                   & -h_x/\sqrt{2}    & {\bf A}_{m B}
\end{array}
\right),
\nonumber
\end{eqnarray}
with
$
{\bf A}_{m R } 
= 
{\bf A}_{m G} 
= 
{\bf A}_{m B} 
= -2t \cos(k_y a - 2\pi m \alpha) ,
$
where the parameter $\alpha = \Phi/\Phi_0$ represents the ratio
between the magnetic flux through a lattice plaquette $\Phi = H a^2$ and
the flux quantum $\Phi_0 = hc/e$, 
or the ratio between the 
plaquette area $a^2$ and the square of the magnetic length 
$\ell_{M} = \sqrt{hc/eH}$, that is, $\alpha = (a/\ell_M)^2$.
The matrix that contains the color-orbit coupling is 
\begin{eqnarray}
{\bf B}
=
\left(
\begin{array}{ccc}
-t e^{-ik_{T}a} & 0    & 0  \\
0                & -t  & 0  \\
0                & 0     & -t e^{ik_{T}a}
\end{array}
\right),
\nonumber
\end{eqnarray}
where $k_T$ $(-k_T)$ corresponds to the momentum transfer along the 
$x$ direction for state $R$ $(B)$, while the momentum transfer for state
$G$ is zero.

\begin{figure}[tb]
\centering 
\epsfig{file=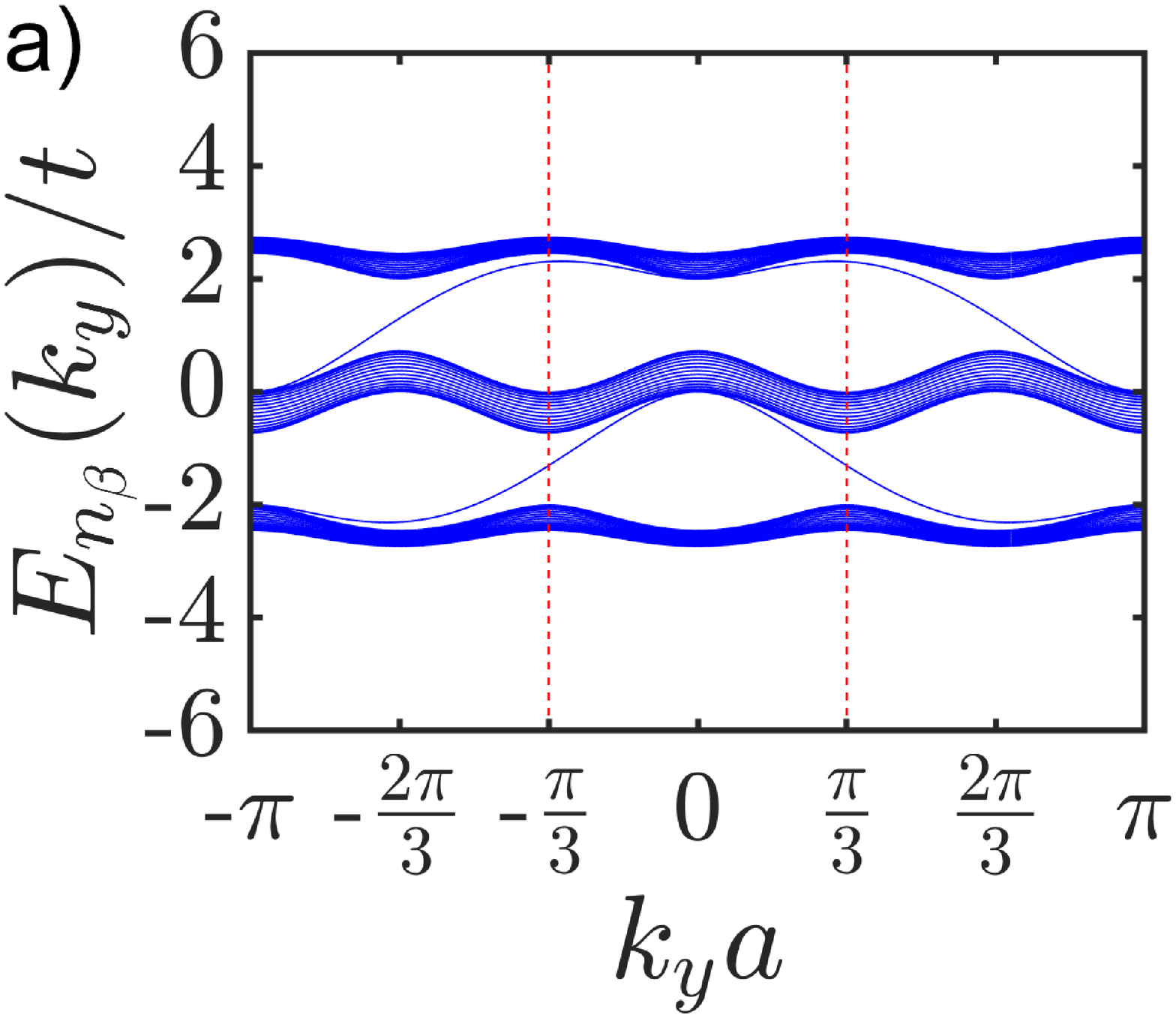,width=0.49 \linewidth}
\epsfig{file=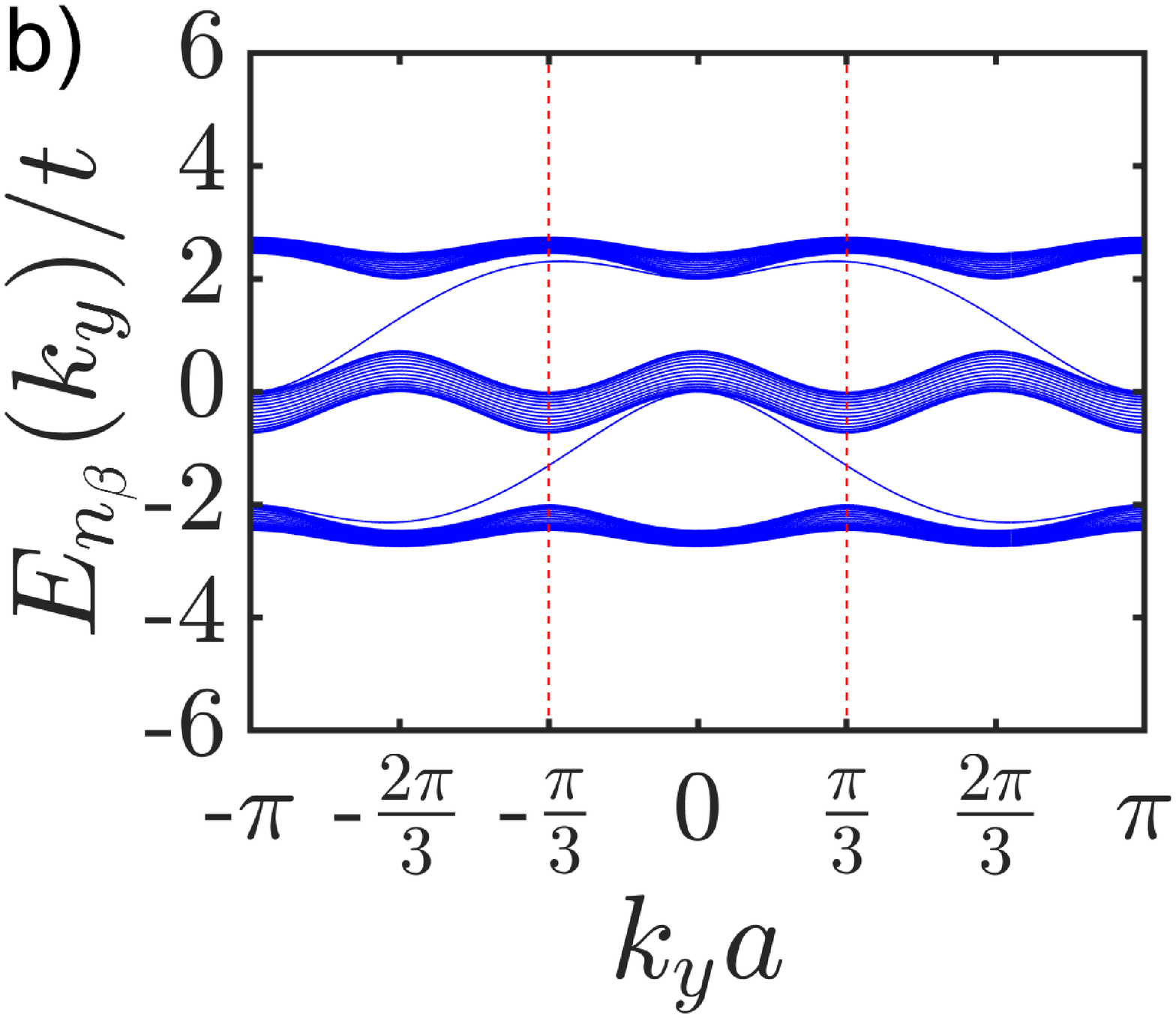,width=0.49 \linewidth}
\vskip 0.2cm
\epsfig{file=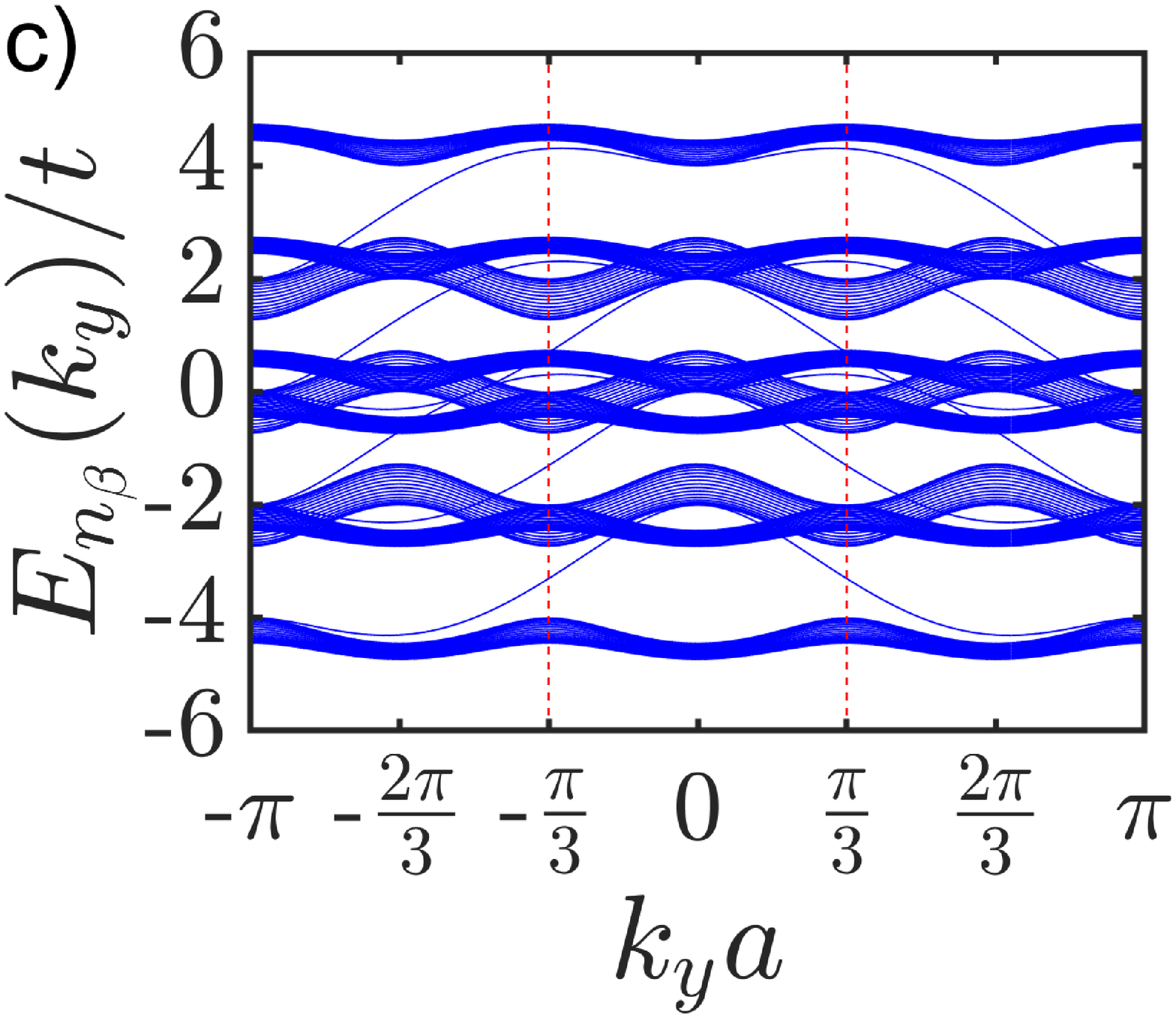,width=0.49 \linewidth}
\epsfig{file=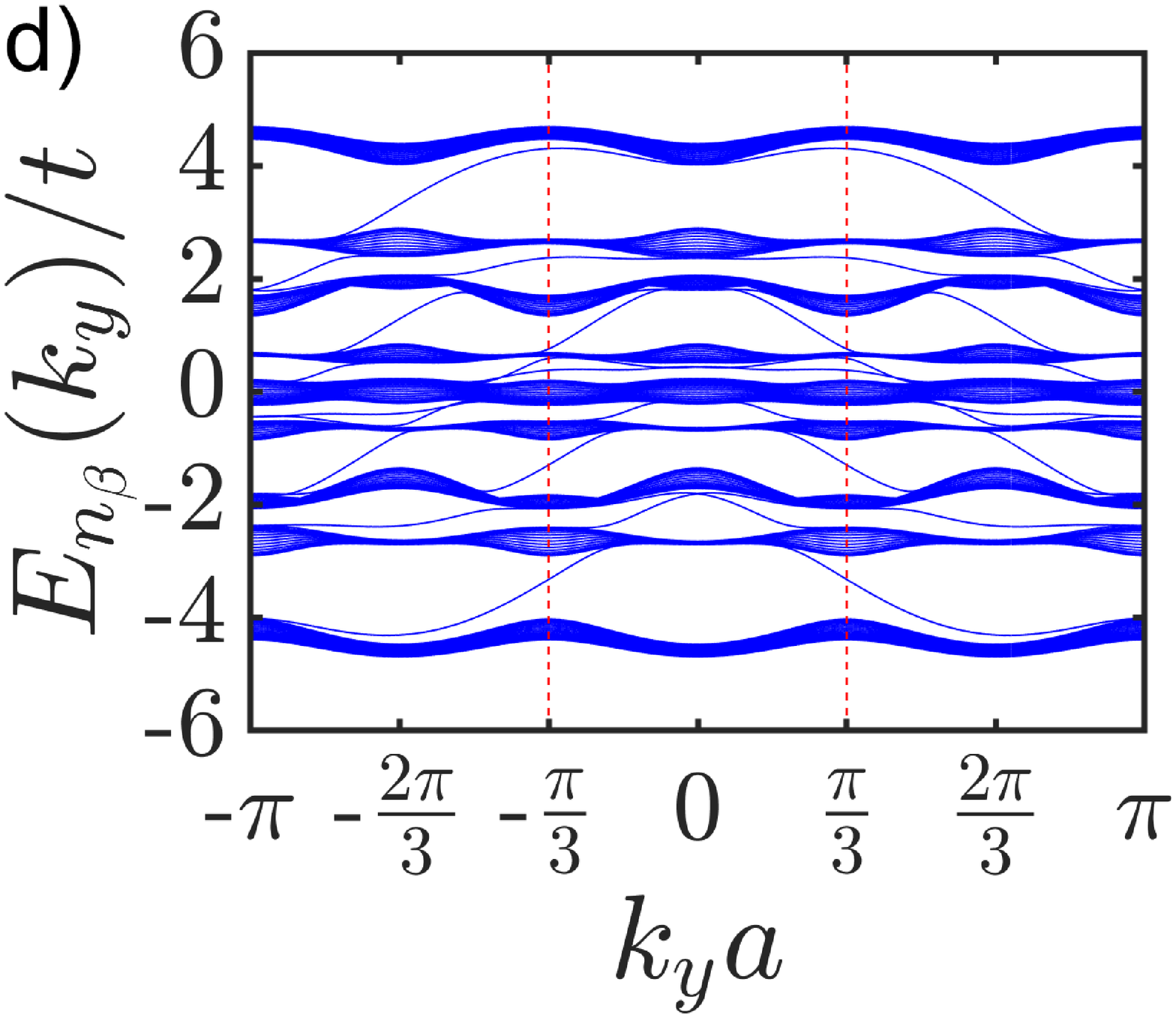,width=0.49 \linewidth}
\caption{ 
\label{fig:one}
(Color Online)
Eigenvalues $E_{n_{\beta}} (k_y)/t$ of the color-dependent Harper's matrix
versus $k_y a$ for magnetic flux ratio $\alpha = 1/3$. 
The parameters are: a) $k_T a = 0$ and $h_x/t = 0$,  
b) $k_T a = \pi/8$ and $h_x/t = 0$, 
c) $k_T a = 0$ and $h_x/t = 2$, d) $k_T a  = \pi/8$ and $h_x/t = 2$. 
The vertical dashed lines located at $k_y a = \pm \pi/3$ indicate the 
boundaries of the magnetic Brillouin zone. The bulk 
bands have periodicity $2\pi/3a$, and the midgap
edge bands have periodicity $2\pi/a$ along the $k_y$ direction.
}
\end{figure}

The full Hofstadter spectrum~\cite{hofstadter-1976} of energy $E$ versus flux
ratio $\alpha = \Phi/\Phi_0$ for colored fermions can be obtained from the
eigenvalues of the Harper's matrix defined
in Eq.~(\ref{eqn:hamiltonian-matrix-cylinder-geometry}). However, in this work,
we focus on a fixed value of $\alpha$ and discuss the energy spectrum as a function of
the color-orbit coupling $k_T$ and color-flip field $h_x$. We consider 
$N = 50$ sites along the $x$ direction, with three states $\{R, G, B\}$ per site, 
but periodic boundary conditions along the $y$ direction. 
The eigenvalues $E_{n_{\beta}}( k_y )$ are labeled by a discrete band 
index $n_{\beta}$ and by momentum $k_y$, and are functions of the color-orbit coupling 
$k_T$, color-flip field $h_x$ and flux ratio $\alpha = \Phi/\Phi_0$.  The index $\beta$ in
$n_{\beta}$ is a reminder that the resulting bands carry a mixed-color index $\beta$, when
color is conserved, the index $\beta$ labels $\{R, G, B\}$ states.

In Fig.~\ref{fig:one}, we show $E_{n_{\beta}} (k_y)$ for flux ratio $\alpha = 1/3$  
in the cases: 
a) $k_T a = 0$ and $h_x/t = 0$, where there are three sets of 
degenerate bulk bands connected by color-degenerate midgap edge bands; 
b) $k_T a = \pi/8$ and $h_x/t = 0$, which is identical to 
case a) because of a color-gauge symmetry that allows gauging away 
the color-orbit coupling; 
c) $k_T a = 0$ and $h_x/t = 2$, where there are nine sets 
of bulk bands with regions of overlap (because 
color-degeneracies are only partially lifted by the color-flip field), 
and where there are color-dependent midgap edge bands connecting bulk bands;
d) $k_T a = \pi/8$ and $h_x/t = 2$, where there are nine sets 
of bulk bands connected by color-dependent midgap edge states, 
but residual bulk band overlaps are lifted by the additional
presence of color-orbit coupling. 
All bulk bands have momentum space periodicity of $2\pi/3a$, 
while all edge bands have period $2\pi/a$ along the $k_y$ direction.
It is important to point out that there are potential experimental techniques to image
directly edge states~\cite{goldman-2013b} in the context of ultracold atoms.
The periodicity of the bulk states is determined by the denominator 
$q$ of the rational magnetic flux ratio $\alpha = p/q$, which 
for $\alpha = 1/3$ corresponds to $q = 3$. 
In Fig.~\ref{fig:one}, the vertical dashed lines 
specify the boundaries of the magnetic Brillouin zone at 
$k_y a = \pm \pi/3$.

Now that we have obtained the eigenspectrum of the system and
identified the existence of midgap edge states connecting different mixed-color bands,
we discuss next the associated Chern numbers for the colored fermions.

\section{Color Chern Numbers}
\label{sec:color-chern-numbers}

To identify topologically non-trivial mixed-color bands and extract their Chern indices,
we impose periodic boundary conditions along the $x$ and $y$ directions, and
compactify our cylinder into a torus.
For rational $\alpha = p/q$, we write the color-dependent Harper's Hamiltonian 
as a $3q \times 3q$ matrix 
\begin{eqnarray}
\label{eqn:hamiltonian-matrix-toroidal-geometry}
{\bf H}({k_x, k_y}) 
= 
\left(
\begin{array}{c c c}
    {\bf H}_{RR} & {\bf H}_{RG} & {\bf H}_{RB}\\
    {\bf H}_{GR} & {\bf H}_{GG} & {\bf H}_{GB}\\
    {\bf H}_{BR} & {\bf H}_{BG} & {\bf H}_{BB}
\end{array}
\right)
\end{eqnarray}
in momentum $(k_x, k_y)$ space, 
by taking advantage of the magnetic translation group. We define 
$q \times q$ block matrices ${\bf H}_{cc^\prime}$, where 
$c$ and $c^\prime$ label the three color states $\{ R, G, B\}$.
The color-diagonal $q \times q$ block matrices 
${\bf H}_{cc}$ are 
\begin{eqnarray}
\left(
\begin{array}{c c c c c}
    \Gamma_1 & -te^{i k_{x c}a} & 0 & \dots  & -te^{-i k_{x c}a} \\
   -te^{-i{k}_{x c}a} & \Gamma_2 & -te^{i k_{x c}a} & \dots  & 0 \\
    \vdots & \vdots & \vdots & \ddots & \vdots \\
    -te^{i k_{x c}a} & 0 & \dots & -te^{-i k_{x c} a}  & \Gamma_q
\end{array}
\right),
\nonumber
\end{eqnarray}
where ${k}_{x c} = k_x -\gamma_c k_T$ is the color-dependent momentum 
along the $x$ direction, including the color-dependent momentum transfer
$\gamma_c k_T$, with $\gamma_R = +1$, $\gamma_G = 0$, and 
$\gamma_B = -1$. The kinetic energy terms are 
$\Gamma_m = -2t\cos(k_ya - 2\pi\alpha m)$, where the magnetic flux ratio
is $\alpha = p/q$ and $m$ takes values $(1, ..., q)$. 
The color-off-diagonal $q \times q$ block matrices are 
${\bf H}_{RB} = {\bf H}_{BR} = {\bf 0}$, and ${\bf H}_{RG} = {\bf H}_{GR}
= {\bf H}_{GB} = {\bf H}_{BG} = {\bf H}_{\rm flip}$, where 
\begin{eqnarray}
{\bf H}_{\rm flip}
=
\left(
\begin{array}{c c c c c}
    -h_x/\sqrt{2} & 0 & 0 & 0  & 0 \\
    0 & -h_x/\sqrt{2} & 0 & 0  & 0 \\
    \vdots & \vdots & \vdots & \ddots & \vdots \\
    0 & 0 & \dots & 0  & -h_x/\sqrt{2}
\end{array}
\right),
\end{eqnarray}
describes color flips between $R$ and $G$ states, as well as, 
between $G$ and $B$ states, via the color-flip field $h_x$.

Next, we analyze the Chern numbers for different values of color-orbit 
coupling and color-flip fields, but fixed flux ratio $\alpha = p/q$. 
The energy
spectrum associated with the Hamiltonian ${\bf H} (k_x, k_y)$ in
Eq.~(\ref{eqn:hamiltonian-matrix-toroidal-geometry}) has $3q$ 
color-magnetic bands $E_{\ell_\gamma} ({\bf k})$ 
that are labeled by a magnetic band number $\ell_\gamma$ with generalized color
index $\gamma$ corresponding to mixed color states, which we identify
as Cyan $(C)$, Magenta $(M)$ and Yellow $(Y)$ or via a pseudo-spin 
basis $\{ C, M, Y \} \to \{ \Uparrow, 0, \Downarrow \}.$ 
The minimum number of gaps between bulk bands is $q-1$, when the bands 
are triply degenerate and the maximum is $3q - 1$, 
when there is no overlap between the bands.

The Chern index for the $\ell_\gamma^{th}$ band with generalized color 
index $\gamma$ 
is
\begin{equation}
\label{eqn:chern-index}
C_{\ell_\gamma}  
= 
\frac{1}{2\pi i}
\int_{\partial \Omega} 
d^2 {\bf k} 
F_{xy}^{(\ell_\gamma)} ({\bf k}),
\end{equation}
where the domain of integration $\partial \Omega$ in momentum space
corresponds to the magnetic Brillouin zone, that is, $\partial \Omega_x = [-\pi/a, \pi/a]$
along the $k_x$ direction,  and $\partial \Omega_y = [-\pi/qa, \pi/qa]$ along
the $k_y$ direction. The function
\begin{equation}
F_{xy}^{(\ell_\gamma)} ({\bf k}) 
= 
\partial_x A_y^{(\ell_\gamma)} ({\bf k}) 
- 
\partial_y A_x^{(\ell_\gamma)} ({\bf k}),
\end{equation}
is the Berry curvature expressed in terms of the Berry connection
$
A_{j}^{(\ell_\gamma)}({\bf k}) 
= 
\langle u_{\ell_\gamma} ({\bf k}) \vert 
\partial_j
\vert u_{\ell_\gamma} ({\bf k}) \rangle
$ 
where $\vert u_{\ell_\gamma} ({\bf k}) \rangle$ are
the eigenstates of the Hamiltonian ${\bf H} (k_x, k_y)$ 
defined in Eq.~(\ref{eqn:hamiltonian-matrix-toroidal-geometry}).
In the limit of zero color-orbit coupling $(k_T = 0)$ and zero 
color-flip field $(h_x = 0)$, the energy spectrum for flux ratio 
$\alpha = p/q$ has triply-degenerate $q$ magnetic bands and $q-1$ gaps, such 
that the Chern index from Eq.~(\ref{eqn:chern-index}) acquires 
a similar form to that found in the quantum Hall effect 
literature for spin-1/2 systems~\cite{thouless-1982, kohmoto-1985}. 

Chern indices are properties of bands  
$E_{\ell_\gamma} ({\bf k})$ or band bundles with degeneracy $D$, and
are computed using a discretized version of Eq.~(\ref{eqn:chern-index})
via a generalization of the method used for spin-$1/2$ systems~\cite{hatsugai-2005}.
However, Chern numbers are defined within band gaps and depend 
on which gap the chemical potential is located.
If the chemical potential $\mu$ is located in a band gap labelled by index $r$
and corresponding to filling factor $\nu = r/q$, then the Chern number 
at this value of $\mu$ is
\begin{equation}
\label{eqn:chern-number}
C_r = \sum_{\ell_\gamma, E < \mu}^{\nu = r/q} C_{\ell_\gamma},
\end{equation}
that is, the sum of Chern indices of bands with 
energies $E < \mu$, which characterize the insulating state labelled by the gap index $r$
and filling factor $\nu = r/q$. Using our normalization, the maximum filling factor is $\nu_{max} = 3$.
Furthermore, via the bulk-edge correspondence~\cite{hatsugai-1996},
the Chern number $C_r$ calculated from the toroidal geometry 
(bulk system without edges) measures the total chirality of 
midgap edge states that are present in the 
cylindrical geometry. As described next, we use the Chern numbers defined in
Eq.~(\ref{eqn:chern-number}) to classify emergent topological phases in
the phase diagrams of chemical potential $\mu$ versus color-flip field $h_x$
and chemical potential $\mu$ versus color-orbit coupling $k_T$.

\section{Phase Diagrams of Color Fermions}
\label{sec:phase-diagrams}

Since we are interested in the effects of color-orbit coupling $k_T$ and color-flip field $h_x$,
we focus on phase diagrams for constant flux ratio $\alpha = p/q$, and choose the particular
value of $\alpha = 1/3$, where non-trivial topological properties emerge. We use the
Chern numbers defined in Eq.~(\ref{eqn:chern-number}) to classify the topological phases in
the phase diagrams of chemical potential $\mu$ versus color-flip field $h_x$ and $\mu$ versus
color-orbit coupling $k_T$. In some situations, a refinement is necessary to distinguish phases with
the same {\it charge} Chern numbers, and we need to monitor the properties of the midgap edge
states to expand the topological classification.

\subsection{Chemical potential versus color-flip field}
\label{sec:chemical-potential-vs-color-flip-field} 

\begin{figure} [tb]
\centering
\epsfig{file=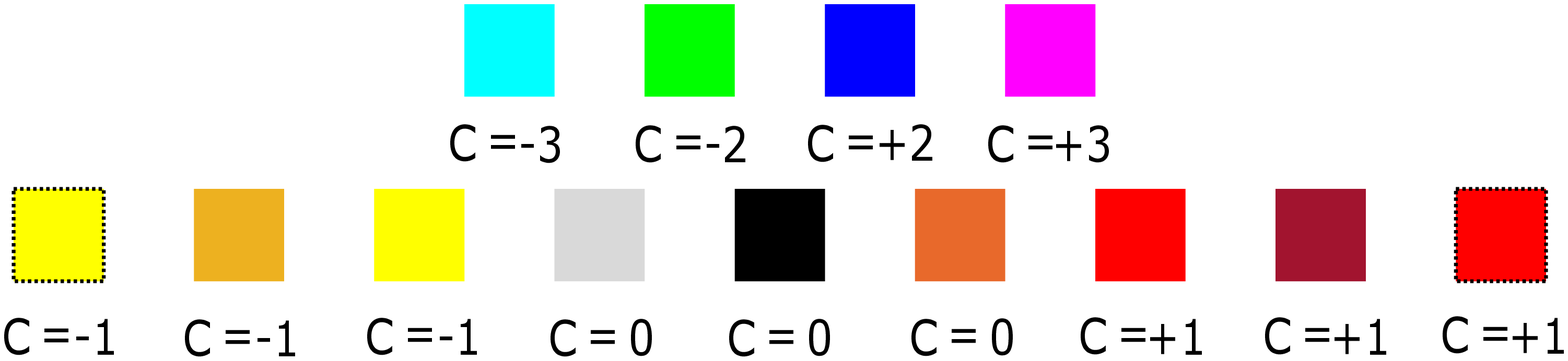,width=0.80 \linewidth}
\vskip 0.2cm
\epsfig{file=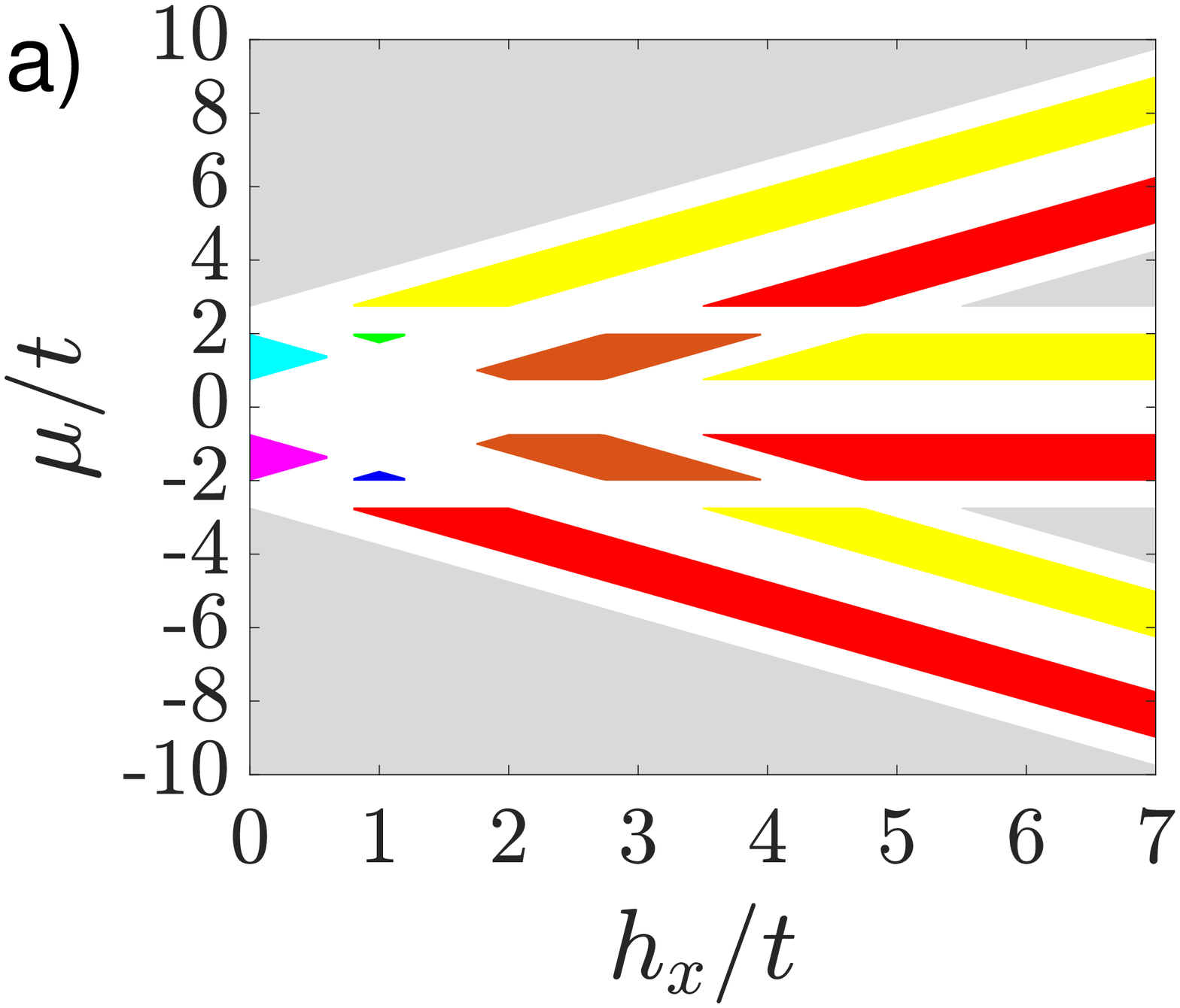,width=0.49 \linewidth}
\epsfig{file=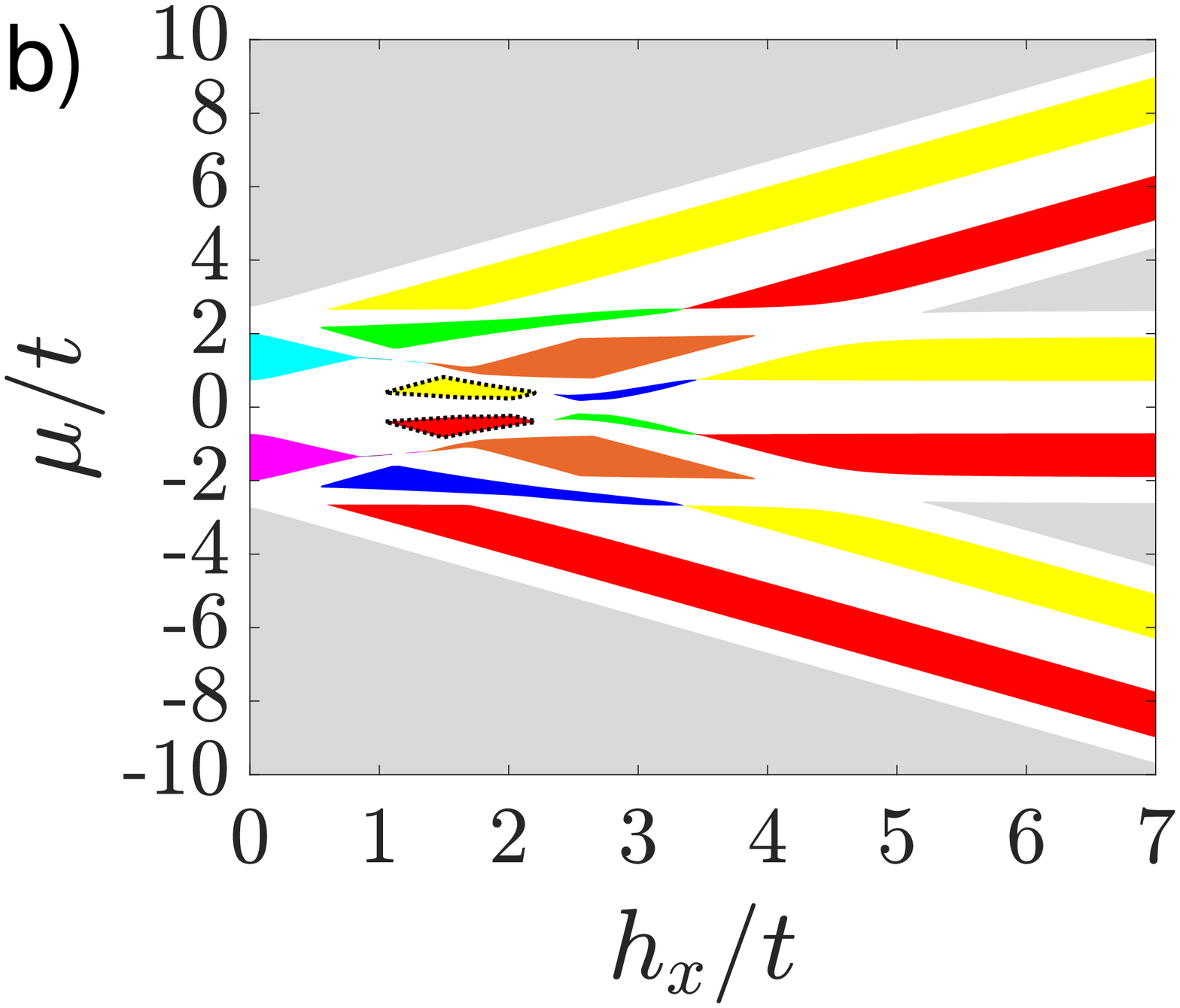,width=0.49 \linewidth}
\vskip 0.2cm
\epsfig{file=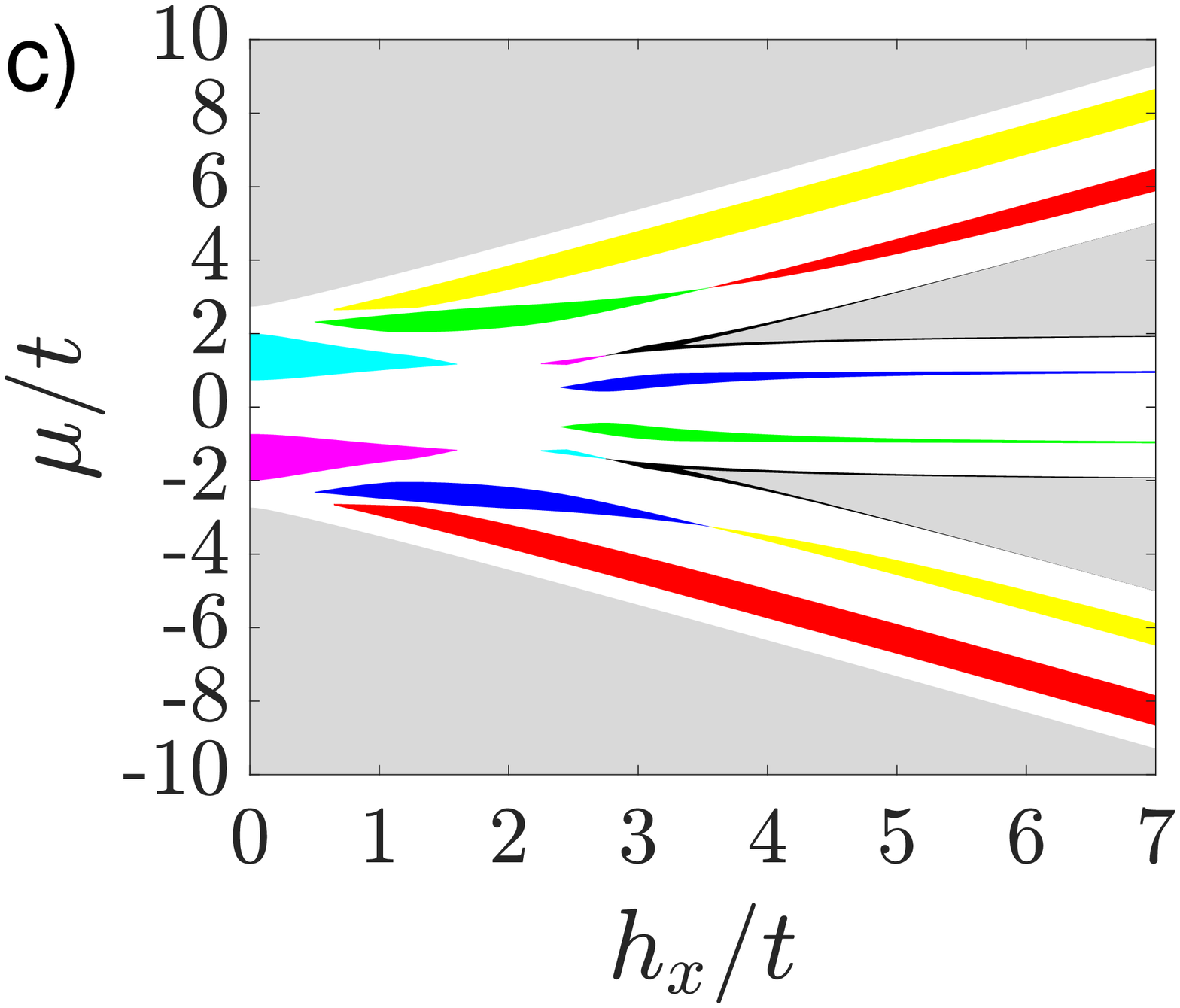,width=0.49 \linewidth}
\epsfig{file=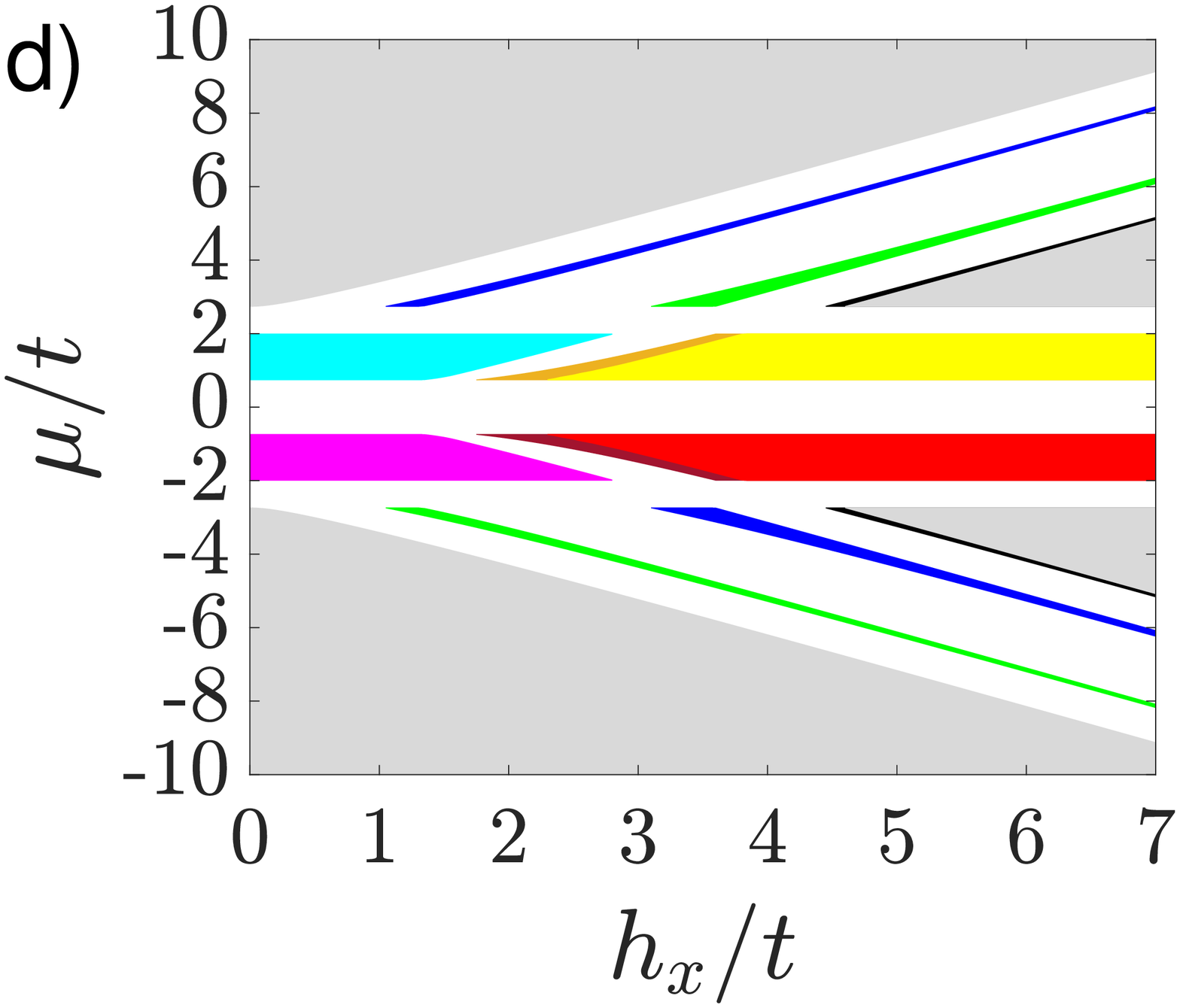,width=0.49 \linewidth}
\caption{ 
\label{fig:two}
(Color Online)
Phase diagrams of chemical potential $\mu/t$ versus
Zeeman field $h_x/t$ and the associated Chern numbers
are shown for spin-orbit coupling parameters: a) $k_T a = 0,$
b) $k_T a = \pi/8,$  c) $k_T a = \pi/2,$ and
d) $k_T a = \pi$. 
The white regions correspond to 
gapless ({\it conducting}) phases, where the chemical potential lies within a band of states,
while the non-white regions correspond to insulating phases, where the chemical potential
lies within the gaps between bands of states. As shown in the color palette (legend), the
Chern numbers for each colored region are 
$+3$ (magenta), $+2$ (blue), $+1$ (red, red-with-black-dots and dark red), 
0 (gray, black, orange), $-1$ (yellow, yellow-with-black-dots and dark yellow), $-2$ (green),
and $-3$ (cyan).
}
\end{figure}

In Fig.~\ref{fig:two}, we show phase diagrams of chemical potential
$\mu$ versus the color-flip field $h_x$ for fixed value of the 
magnetic flux ratio $\alpha = 1/3$ with four values of 
the color-orbit parameter:
a) $k_T a = 0$, b) $k_T a = \pi/8$, c) $k_T a = \pi/2$, and 
d) $k_T a = \pi$. From the figures, it is clear that phase diagrams are quite complex,
in particular for values of $h_x/t > 1.$ But, before we embark on the description of the
phase diagrams for each figure, we discuss first the labelling of the regions indicated
in the legend of the figure.

In Fig.~\ref{fig:two}, the white regions correspond to 
gapless ({\it conducting}) phases, where the chemical potential lies within a band of states,
while the non-white (colored) regions correspond to insulating phases, where the
chemical potential lies within the gaps between bands of states. The legend in this figure
is a color palette describing the Chern numbers for each colored region: $+3$ (magenta), 
$+2$ (blue), $+1$ (red, red-with-black-dots and dark red), 
$0$ (gray, black, orange), $-1$ (yellow, yellow with black dots and dark yellow), $-2$ (green),
$-3$ (cyan). There are also very small regions with Chern numbers $\pm 6$ only seen
in Fig.~\ref{fig:two}b. Given that certain regions with different colors have the same
Chern numbers, it is clear that additional properties are needed to distinguish them.
Before we discuss the phase diagrams in more detail, we describe first the color palette.
The magenta (cyan) regions with Chern number $+3$ $(-3)$ possess three chiral midgap
edge states with positive (negative) chirality,
while the blue (green) regions with Chern number $+2$ $(-2)$ possess two chiral midgap edge
states with positive (negative) chirality.
The red (yellow) regions with Chern number $+1$ $(-1)$ possess
one chiral midgap edge state with positive (negative) chirality;
the dark red (dark yellow) regions with Chern number $+1$ $(-1)$ possess not only
one chiral midgap edge state with positive (negative) chirality, but also present achiral
midgap edge states, nevertheless they are not topologically distinct from their parent red (yellow) regions.
However, the red-with-black-dots (yellow-with-black-dots) regions with
Chern number $+1$ $(-1)$ have three chiral midgap edge states, two of which have positive
(negative) chirality, and one of which has negative (positive) chirality. Therefore, these insulating
regions are topologically distinct from the red (yellow) regions with the same Chern numbers.
The gray regions with Chern number $0$ are topologically trivial with no chiral or achiral midgap
edge states. The black regions with Chern number $0$ are also topologically trivial with no chiral
midgap edge states but with achiral midgap edge states. These black regions are not
topologically distinct from the gray regions. Lastly, the orange regions with Chern number $0$
are topologically non-trivial and possess two chiral midgap edge states with opposite chirality,
reminiscent of the quantum spin-Hall effect in spin-$1/2$ fermions.

Before we analyze the details of the phase diagrams shown in Figs.~\ref{fig:two}a through
\ref{fig:two}d, we discuss first some of their general properties. The first thing to
notice is that the phase diagrams are particle-hole symmetric with respect to the line $\mu =0$,
this means that the Chern numbers have odd symmetry upon reflection through $\mu = 0$, since they
represent the total chirality of edge states. A change in the {\it charge} from particle-like
to hole-like leads to a flip in the chirality of the edge states, that is, a change in sign of the Chern number
defined in Eq.~(\ref{eqn:chern-number}). The particle-hole symmetry in the Chern number spectrum
reflects the same symmetry present in the Hamiltonian for this problem.

For fixed flux ratio $\alpha = 1/3$ and color-orbit parameter $k_T a$, the number of insulating phases,
where gaps between energy bands exist, grows from two to eight with increasing color-flip
parameter $h_x/t$ from $h_x/t \ll 1$ to $h_x/t \gg 1.$ This situation does not occur 
in electronic systems with spin-$1/2$ since the Zeeman field $h_x$ cannot be tuned 
independently from the magnetic ratio $\alpha$ as they have the same origin, 
and typically $h_x$ has very small values in comparison to 
the hoping parameter $t$, such that $h_x/t \ll 1$. 
However, for ultracold fermions, since $h_x$ is a 
synthetic field that can be tuned independently from the magnetic ratio 
$\alpha$, it can attain high values in comparison to $t$ and provide access to phases that are not
encountered in standard condensed matter systems.

In the regime $h_x/t \ll 1$, where the color splitting caused by
the color-flip field $h_x$ is small in comparison to the hopping $t$, the energy spectrum has
only two gaps similar to the cases illustrated in Figs.~\ref{fig:one}a and~\ref{fig:one}b.
Thus, only two insulating phases emerge: one
with Chern number $+3$ (magenta region) at the first gap and
the other $-3$ (cyan region) at the second
gap. These phases are the color generalizations of the quantum Hall phases for spin-$1/2$ systems,
which have Chern numbers $+2$ and $ -2$.
In the phase with Chern number $+3$, there are
three chiral midgap edge states with positive chirality, and
in the phase with Chern number $-3$ there are three chiral midgap
edge states with negative chirality. Hence, these are essentially color unpolarized phases.

In the regime $h_x/t \gg 1$, where the color splitting caused by
the color-flip field $h_x$ is large in comparison to the hopping $t$, in which case the system is essentially
polarized in a mixed-color basis of the color (pseudo-spin) matrix ${\bf J}_x$
described in Eq.~(\ref{eqn:hamiltonian-matrix-spin-1}).
In this case, the color-orbit parameter $k_T a$ lifts band degeneracies and creates
eight gapped phases. In Figs. ~\ref{fig:two}a and~\ref{fig:two}b where
$k_T a \ll \pi$, the eight color-insulating phases have  Chern numbers $+1$ (red regions),
$-1$ (yellow regions), and $0$ (gray regions), when $h_x/t \gg 1$. However,
when color-orbit coupling $k_T a$ is sufficiently large, the mixed color-bands get strongly
coupled and the nature of the insulating phases changes dramatically.
For $k_T a = \pi/2$, the insulating states have
Chern numbers $+1$ (red regions), $-1$ (yellow regions), $0$ (gray regions),
$-2$ (green region) and $+2$ (blue region); while for $k_T a = \pi$, the insulating phases
have $-2$ (green regions), $+2$ (blue regions), $0$ (gray regions),
$+1$ (red region), and $-1$ (yellow region). 

In Fig.~\ref{fig:two}a, where $k_T a = 0$, we can see many different insulating phases in the chemical
potential versus color-flip field diagram as $h_x/t$ is changed. Each of these insulating phases corresponds
to a fixed filling factor $\nu = r/3$, where $r$ is an integer that represents a gap, varying from $0$ to $9$.
Zero filling factor $\nu = 0$ corresponds to $r = 0$, and full filling factor $\nu = 3$ corresponds to $r = 9$,
these correspond to the lower and upper gray regions respectively,
which are topologically trivial with zero Chern number.
Viewing the $\mu/t$ versus $h_x/t$ phase diagram through the lens of filling factors $\nu$ reveals that 
there is one insulating phase with filling factor $\nu = 1/3$ (lower red region),
two insulating phases with $\nu = 2/3$ (blue and lower yellow regions),
three insulating phases with $\nu = 1$ (magenta, lower orange and lower gray regions),
one insulating phase with $\nu = 4/3$ (middle red region),
one insulating phase with $\nu = 5/3$ (middle yellow region),
three insulating phases with $\nu = 2$ (cyan, upper orange and upper gray regions),
two insulating phases with $\nu = 7/3$ (green and upper red regions),
and one insulating phase with $\nu = 8/3$ (upper yellow region).

The most interesting feature in Fig.~\ref{fig:two}a are the orange regions around $\mu = 0$
with filling factors $\nu = 1$ and $2$, which possess zero Chern number, but have two chiral midgap edge
states with opposite chiralities, producing a phase that we name
quantum color Hall (QCoH) insulator, in analogy to the quantum spin Hall (QSH) insulator
that exhibits the quantum spin-Hall effect. These orange regions are the
color versions of the QSH phases for spin-$1/2$ fermions~\cite{kane-2005, haldane-2005, zhang-2006}.

The phase diagram in Fig.~\ref{fig:two}b, where $\alpha = 1/3$ and  $k_T a = \pi/8$, has the
most unconventional phases and phase transitions. The insulating phases for fixed filling factor
are as follows.
At filling factors $\nu = 0$ and $\nu = 3$, the bands are either completely empty of completelly full,
leading to the topological trivial insulating gray regions.
At filling factor $\nu = 1/3$, there is one insulating phase (lower red region).
At filling factor $\nu = 2/3$, there are two insulating phases (blue and yellow regions)
and a direct topological quantum phase transition
between blue and yellow regions at $h_x/t = 3.35$, and by particle-hole symmetry,
there are also, at filling factor $\nu = 7/3$,  two insultating phases (green and red regions) and
a direct topological phase transition between the green and red regions at the same color-flip field
$h_x/t = 3.35$.
At filling factor $\nu = 1$, there are five insulating phases.
Starting from the magenta region, as $h_x/t$ grows, 
there is a direct topological quantum phase transition at $h_x/t = 0.85$ to a
very thin and small dark magenta region with Chern number +6,
which exists between $0.85 < h_x/t < 1.10$,
and leads to a {\it conducting} phase between $1.10 < h_x/t < 1.20$.
Then another very thin and small magenta region emerges between $1.20 < h_x/t < 1.35$, leading
to an additional direct topological quantum phase transition into the orange region at $h_x/t = 1.35$.
Finally, at high values of $h_x/t$, a topologically trivial gray region emerges.

By particle-hole symmetry, at filling factor $\nu = 2$ there are also five insulating phases, which
have Chern numbers opposite to the ones for $\nu = 1$. 
Starting from the cyan region, as $h_x/t$ grows, 
there is direct topological quantum phase transition at $h_x/t = 0.85$ to a
very thin and small dark cyan region with Chern number -6,
which exists between $0.85 < h_x/t < 1.10$,
and leads to a {\it conducting} phase between $1.10 < h_x/t < 1.20$.
Then another very thin and small cyan region emerges between $1.2 < h_x/t < 1.35$, leading
to an additional direct topological quantum phase transition into the orange region at $h_x/t = 1.35$.
Finally, at high values of $h_x/t$, a topologically trivial gray region emerges.

At filling factor $\nu = 4/3$, there are three insulating phases as $h_x/t$ grows
(red-with-black-dots, green and red regions).  There is a direct topological quantum phase transition between
the green and red regions at $h_x/t = 3.45$. The green region with two chiral midgap edge
states with negative chirality and the red region with one chiral midgap edge state
with positive chirality exhibit
the standard quantum Hall effect, which we name from now on as the 
quantum charge Hall (QChH) effect. The red-with-black-dots region contains
three midgap edge states, two with positive chirality and one with negative chirality,
thus possessing the QChH effect, however,
two of the midgap edge states with opposite chirality, have also different mixed-color indices
leading to a quantum color Hall (QCoH) effect. This situation is analogous to the quantum spin Hall effect
for spin-$1/2$ fermions.

Again, by particle-hole symmetry, at filling factor $\nu = 5/3$, there
are three insulating phases with growing $h_x/t$
(yellow-with-black-dots, blue and yellow regions). There is a direct topological quantum phase transition between
the green and red regions at $h_x/t = 3.45$. The blue region with two chiral midgap edge
states with positive chirality and the yellow region with one chiral midgap edge state
with negative chirality exhibit
the standard QChH effect. The yellow-with-black-dots region contains three midgap edge states,
two with negative chirality and one with positive chirality, leading to net negarive chirality and the QChH effect.
Furthermore, since midgap edge states with opposite chirality have different mixed-color indices, this phase
also exhibits a QCoH effect.

In Fig.~\ref{fig:two}c, with $\alpha = 1/3$ and $k_T a  = \pi/2$, the insulating
phases for fixed filling factor $\nu$ are as follows.
At filling factors $\nu = 0$ and $\nu = 3$, the bands are either completely empty or completelly full,
leading to the topological trivial insulating gray regions.
For $\nu = 1/3$, there is only the lower red region, and by particle-hole symmetry,
for $\nu = 8/3$, there is only the upper yellow region.
For $\nu =2/3$, there are two insulating phases (blue and yellow regions) with 
a direct topological quantum phase transition between the two of them at $h_x/t  = 3.55$, and
again by particle-hole symmetry,
for $\nu = 7/3$, there are two insulating phases (green and red regions) with a direct topological quantum phase
transition between them at $h_x/t  = 3.55$.
For $\nu = 1$, there are four insulating phases (magenta, cyan, black and gray regions)  with a direct
topological quantum phase transition between the cyan and black phases at $h_x/t  = 2.75$.
A crossover line between the black and gray regions occurs where achiral midgap edge states
from the black regions merge into the bulk of the gray regions, which possess no midgap edge states.
Again, by particle-hole symmetry, at $\nu =2$ there are four insulating phases
(cyan, magenta, black and gray regions) with a direct
topological quantum phase transition between the magenta and black phases at $h_x/t = 2.75$.
A crossover line between the black and gray regions occurs where the achiral midgap edge states
from the black regionsmerge into the bulk of the gray regions, which possess no midgap edge states.

In Fig.~\ref{fig:two}d, with $\alpha = 1/3$ and $k_T a = \pi$, the insulating phases for
fixed filling factor are as follows.
At filling factors $\nu = 0$ and $\nu = 3$, the bands are either completely empty or completelly full,
leading to the topological trivial insulating gray regions.
For filling factor $\nu = 1/3$, there is one insulating phase (lower green region). 
For $\nu = 2/3$, there is one insulating phase (lower blue region).
For $\nu = 1$, there are three insulating phases (magenta, black and gray regions):
the black and gray regions are separated by a line,
where achiral midgap edge states from the black region merge into the bulk leading
to the gray regions with no midgap edge states.
For $\nu = 4/3$, there are two insulating phases (dark red and red regions) separated by a
crossover line, where achiral midgap edge states from the dark red region merge into the bulk
leading to the red region with a single chiral midgap edge state with positive chirality.

The rest of the phase diagram can be obtained by particle-hole symmetry.
For $\nu = 5/3$,  there are two insulating phases (dark yellow and yellow regions)
separated by a crossover line, where achiral edge states, that exist in the dark yellow region, merge
into the bulk leading to the yellow region with one chiral midgap edge state with negative chirality.
For $\nu = 2$, there are three insulating phases (cyan, black and gray regions).
The black and gray regions are separated by a line where achiral midgap edge states from the
black region merge into the bulk leading to the gray regions with no midgap edge states.
For $\nu = 7/3$, there is one insulating phase (upper green region). 
For $\nu = 8/3$, there is one insulating phase  (upper blue region).
We highlight that the dark red and dark yellow regions have not only one topologically non-trivial
chiral midgap edge band, but also topologically trivial achiral midgap edge states. However the
dark red and dark yellow regions are not topologically distinct from
the red and yellow regions which have only one topologically non-trivial chiral
midgap edge band.

In the analysis of different panels of the chemical potential $\mu/t$ versus color-flip fields $h_x/t$
shown in Fig.~\ref{fig:two}, we have seen that the filling factor $\nu = r/q$ can be used as a label for the
insulating phases, because when the chemical potential $\mu$ varies within a gap,
the filling factor remains constant, as the insulating state is incompressible.
Thus, next, we generalize the gap-labelling theorem
found for spin-$1/2$ fermions for fixed magnetic flux parameter $\alpha$, and extend it to fermions
with three internal states.

\subsection{Gap Labelling Theorem}
\label{sec:gap-labelling-theorem}

For spin-$1/2$ fermions in condensed matter systems, a gap labelling theorem that relates  the filling
factor $\nu= r/q$ and the magnetic flux parameter $\alpha = \Phi/\Phi_0 = p/q$ was found by
Wannier and Claro~\cite{wannier-1978, claro-1979}. In that case, the Zeeman field $h_x$ was
neglected and the theorem covered only couplings to the charge degrees of freedom of the system.
However, in the present case, it is clear that the color-flip fields $h_x$ play an important role in creating additional
gaps, as we have seen in the discussion of Fig.~\ref{fig:two}. As $h_x/t$ varies from $h_x/t \ll 1$ to
$h_x/t \gg 1$ the number of gaps grows from two (2) to eight (8). From the analysis of Fig.~\ref{fig:two},
it is clear that the filling factor $\nu$ associated with gap $r$ is a good label for the insulating phases,
as are the Chern numbers $C_r$.

We can establish a relation between the Chern numbers $C_r$, the magnetic ratio $\alpha = p/q$
and the filling factor $\nu = r/q$ by rewriting the Diophantine equation
\begin{equation}
\label{eqn:diophantine}
r = q S_r + p C_r,
\end{equation}
where the integer index $r$ labels the gaps in the energy spectrum 
$E_{\ell_\gamma} (k_x, k_y)$ of the toroidal geometry,
$C_r$ is the Chern number for the $r^{th}$ gap  and
$S_r$ is a supplementary topological invariant.
This  equation can be rewritten in terms of the filling factor 
$\nu = r/q$ and the magnetic ratio $\alpha = p/q$ as 
\begin{equation}
\label{eqn:gap-labelling}
\nu = S_r + \alpha C_r.
\end{equation}
The relation shown above generalizes 
the gap labeling theorem~\cite{wannier-1978, claro-1979} used in the context of the
integer quantum-Hall effect, because the topological quantum numbers
$(S_r, C_r)$ change not only as a function of the magnetic ratio $\alpha$, 
but also as a function of the color-flip field $h_x/t$ and color-orbit 
parameter $k_T a$, that is, $S_r (h_x/t, k_T a)$ and $C_r (h_x/t, k_T a)$.
Notice that the maximal value of $S_r$ for a given gap labeled by $r$ 
is linked to the minimum value of $C_r$ and vice-versa, that is,
$S_{r, {\rm max}} = \nu - \alpha C_{r, {\rm min}}$ and $S_{r, {\rm min}} = \nu - \alpha C_{r, {\rm max}}$.

For three color states, the integer values of $r$ range from $0$ (when $\nu = 0$)  to $3q$
(when $\nu = 3$). In the phase diagrams of $\mu/t$ versus $h_x/t$ or $\mu/t$ versus $k_T a$,
the locations where the gaps open change as a function of $h_x/t$ and $k_T a$,
but the gap labelling relation given in Eq.~(\ref{eqn:gap-labelling}) applies to all insulating phases. 
Just to mention a couple of examples for $\alpha = 1/3$,  the five insulating phases
with filling factor $\nu = 1$ $(r = 3)$ in Fig.~\ref{fig:two}b have topological labels
$(S_3, C_3)$ that vary as a function of the color-flip field $h_x/t$. The sequence of insulating
phases from low to high $h_x/t$ is magenta with $(0, +3)$, dark magenta $(-1, +6)$,
magenta with $(0, +3)$ , orange with $(+1, 0)$, and gray with $(+1, 0)$. While for
the case of $\nu = 2$ $(r = 6)$ in Fig.~\ref{fig:two}b the sequence of insulating phases
with growing $h_x/t$ is cyan with $(+3, -3)$, dark cyan with $(+4, -6)$, cyan with $(+3, -3)$,
orange with $(+2, 0)$, and gray with $(+2, 0)$.

Having introduced the gap labelling relation that connects filling factors $\nu$, magnetic flux ratio
$\alpha$ and topological numbers $(S_r, C_r)$, we will use this ordered pair to classify the topological
phases in the {\it charge} sector, noting that additional topological numbers may arise
in the {\it color} sector, for instance in the phase diagram of $\mu/t$ versus $k_T a$ discussed next.

\subsection{Chemical potential versus color-orbit coupling}
\label{sec:chemical-potential-vs-color-orbit-coupling}
\begin{figure} [tb]
\centering 
\epsfig{file=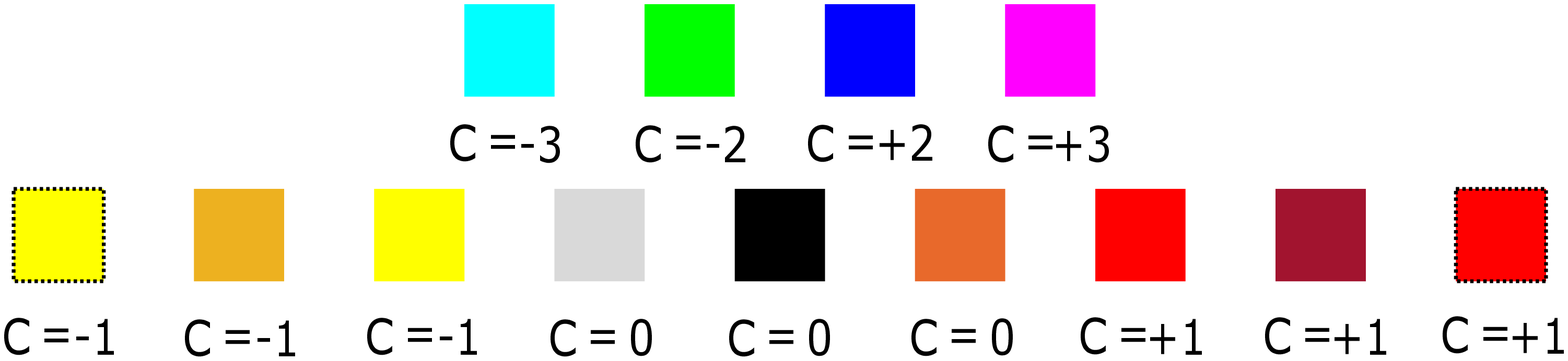,width=0.80\linewidth}
\vskip 0.2cm
\epsfig{file=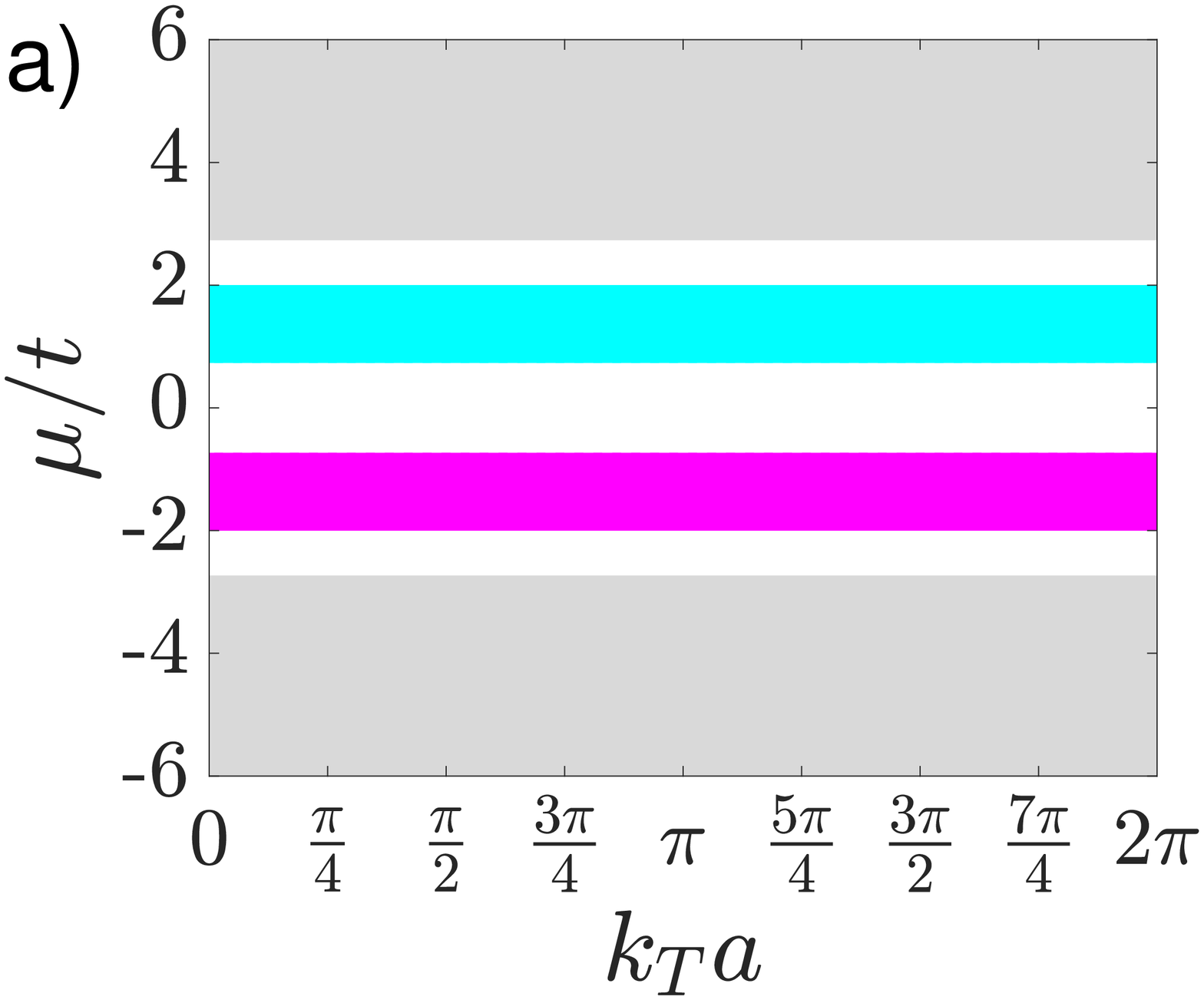,width=0.49 \linewidth}
\epsfig{file=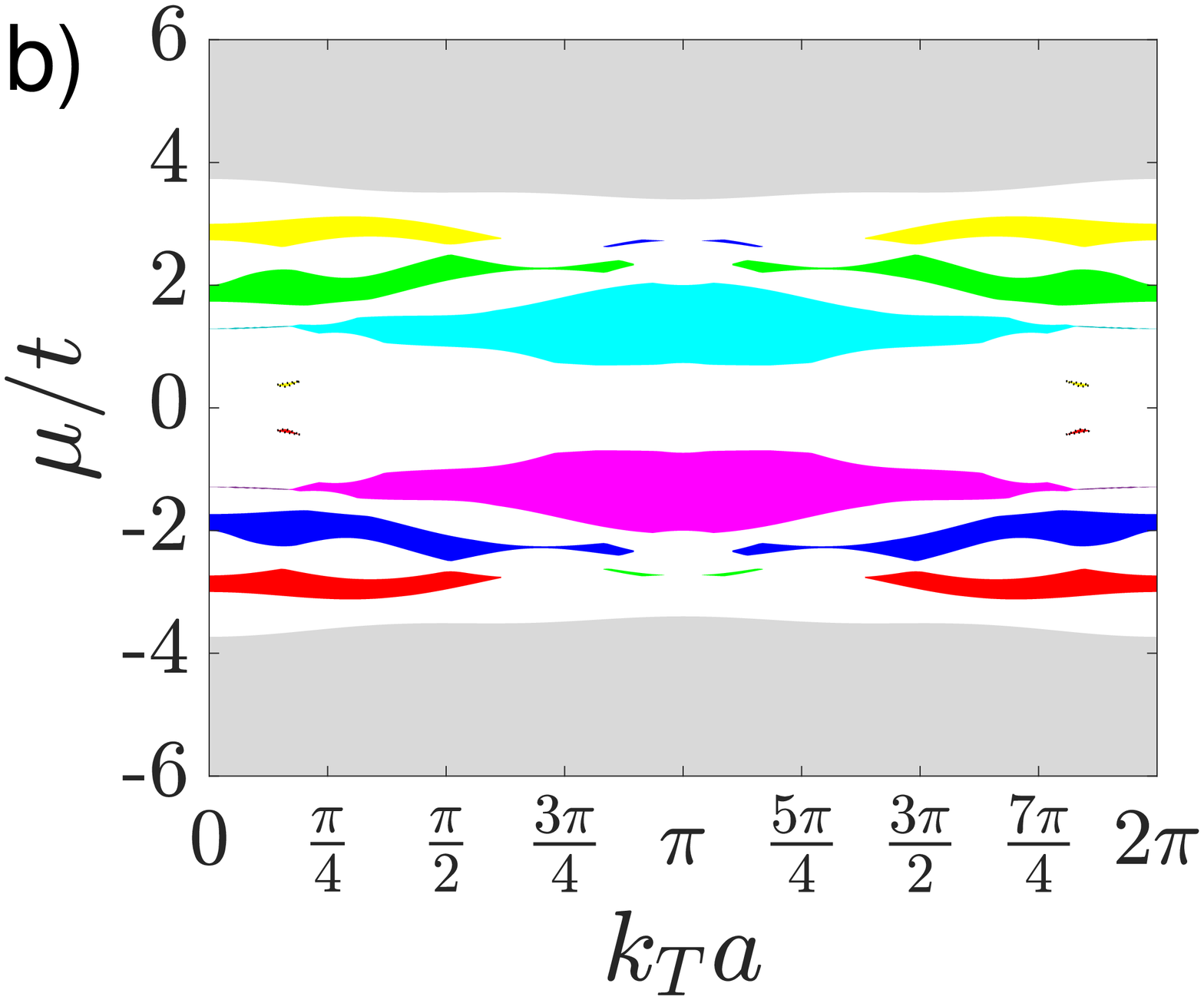,width=0.49 \linewidth}
\vskip 0.2cm
\epsfig{file=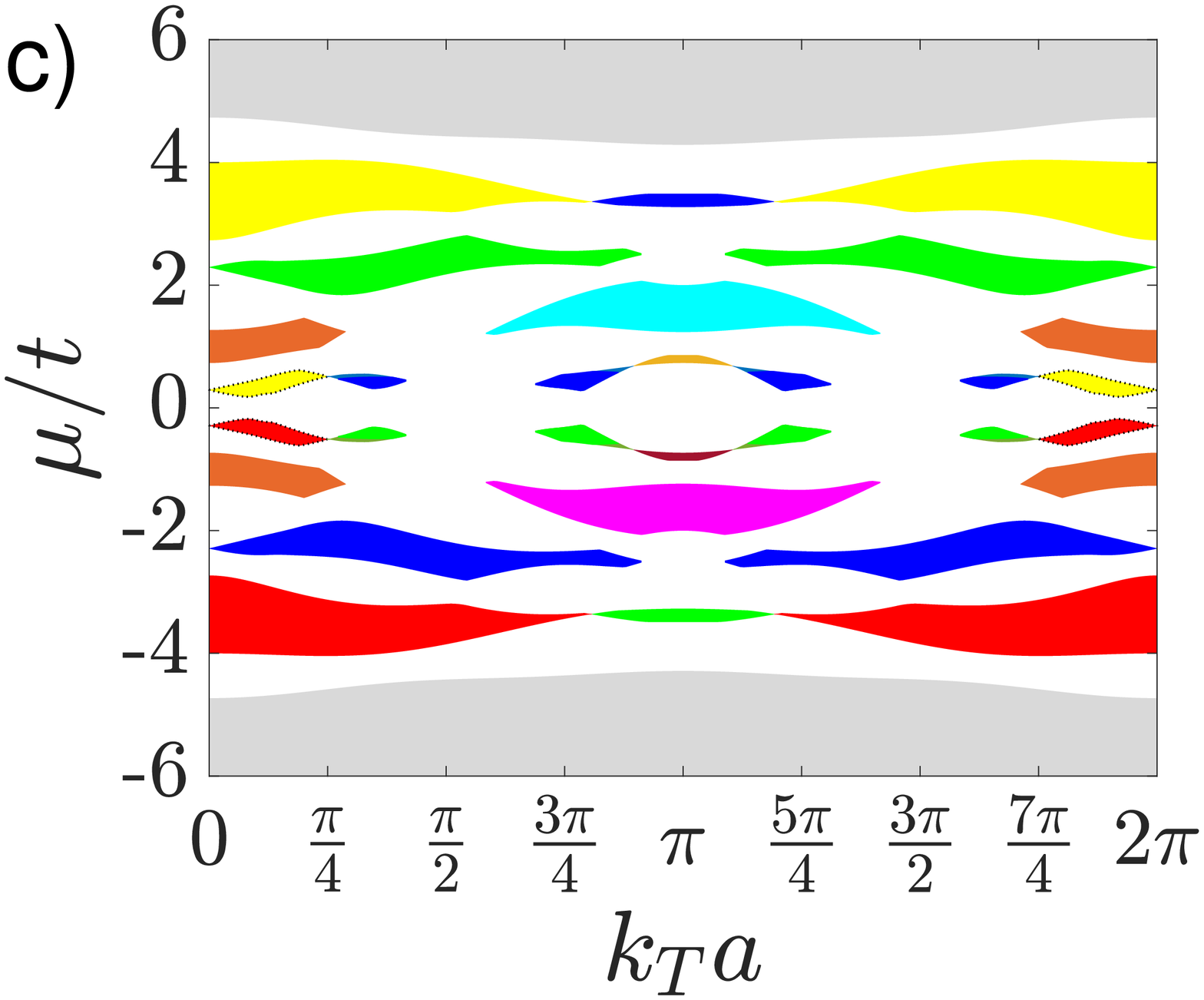,width=0.49 \linewidth}
\epsfig{file=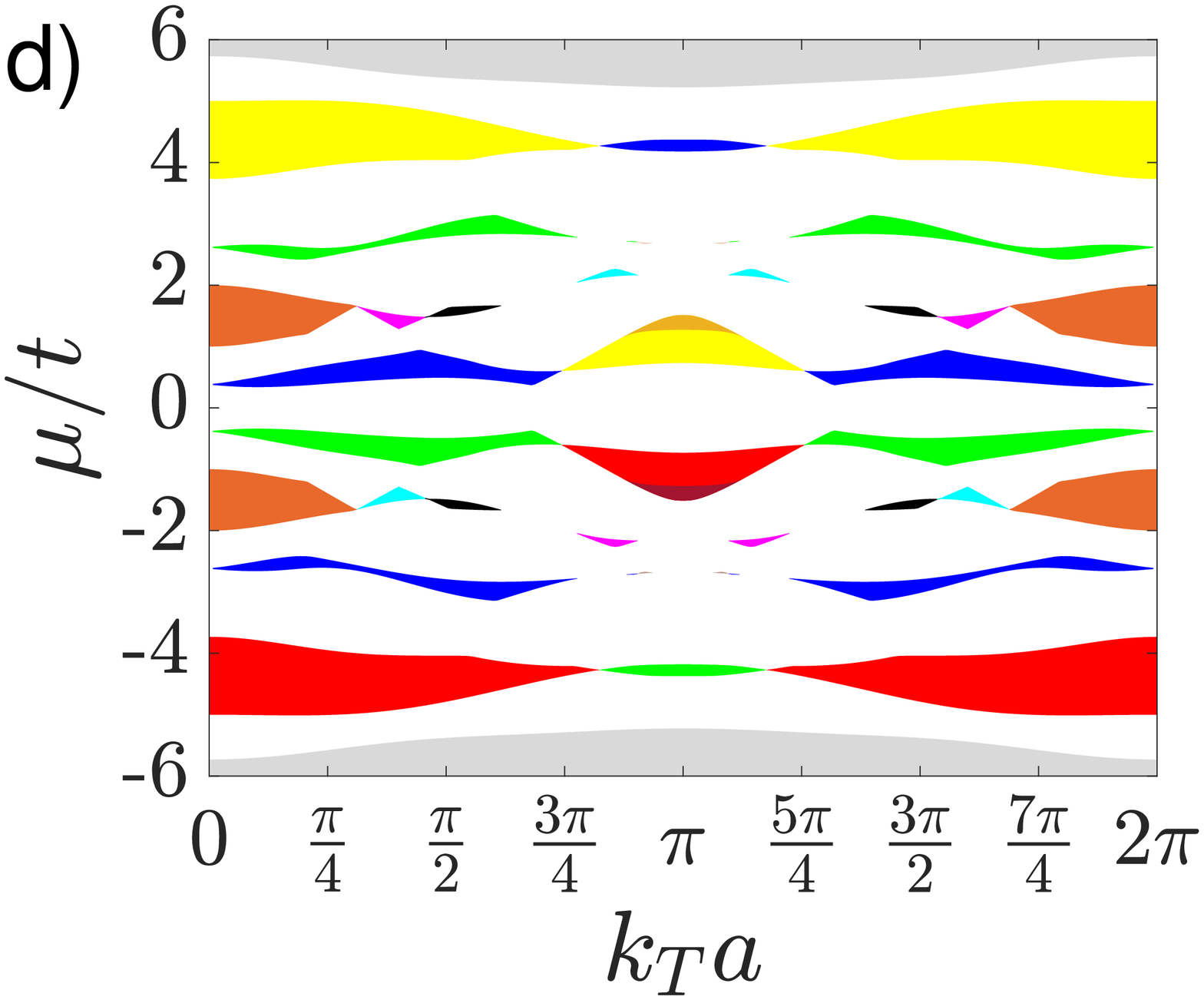,width=0.49 \linewidth}
\caption{ 
\label{fig:three}
(Color Online) 
Chemical potential $\mu/t$ versus spin-orbit parameter $k_T a$ for
flux ratio $\alpha = 1/3$ and Zeeman fields: 
a) $h_x/t = 0$,
b) $h_x/t = 1$,
c) $h_x/t = 2$,
d) $h_x/t = 3$.
The color palette for insulating phases is the same used in 
Fig.~\ref{fig:two} with six additional very small regions with
Chern numbers $+6$ (dark magenta) at $\nu =1$ and $-6$ (dark cyan)
at $\nu = 2$ in Fig.~\ref{fig:three}b;
Chern numbers $+2$ (dark green) at $\nu = 4/3$ and $-2$ (light blue) at $\nu = 5/3$
in Fig.~\ref{fig:three}c; and
Chern numbers $+5$ (light brown) at $\nu = 2/3$ and $-5$ (light-pink) at $\nu = 7/3$
in Fig.~\ref{fig:three}d.
}
\end{figure}

In Fig.~\ref{fig:three}, we show phase diagrams of chemical 
potential $\mu/t$ versus spin-orbit parameter $k_T a$ 
for fixed magnetic ratio $\alpha = 1/3$ and changing color-flip fields: 
a) $h_x/t = 0$, b) $h_x/t = 1$, c) $h_x/t = 2$ and d) $h_x/t = 3$. 
The color palette for insulating phases is the same used in 
Fig.~\ref{fig:two} with six additional very small regions with
Chern numbers $+6$ (dark magenta) at $\nu =1$ and $-6$ (dark cyan)
at $\nu = 2$ in Fig.~\ref{fig:three}b;
Chern numbers $+2$ (dark green) at $\nu = 4/3$ and $-2$ (light blue) at $\nu = 5/3$
in Fig.~\ref{fig:three}c; and
Chern numbers $+5$ (light brown) at $\nu = 2/3$ and $-5$ (light-pink) at $\nu = 7/3$
in Fig.~\ref{fig:three}d.
Before we go into the details of each panel of Fig.~\ref{fig:three}, we point out a few general
properties. We notice that the phase diagram of $\mu/t$ versu $k_T a$ has periodicity of $2\pi$,
inversion symmetry with respect to $k_T a = \pi$, and particle-hole symmetry with
respect to $\mu = 0$. All these properties arise directly from symmetries of the Hamiltonian of the system
discussed in Sec.~\ref{sec:hamiltonian}. Furthermore, for fixed $h_x/t$, there are several
topological quantum phase transitions that occur between different insulating phases as $k_T a$ is changed. 
Notice also that the lower and upper gray regions are topologically trivial  and correspond to
$\nu = 0$ with $(S_0, C_0) = (0,0)$, and
$\nu = 3$ with $(S_9, C_9) = (+3,0)$, respectivelly.

In Fig.~\ref{fig:three}a, where $\alpha = 1/3$ and $h_x/t = 0$, the phase diagram
of $\mu/t$ versus $k_T a$ has only two topological insulating phases. The first one is the 
magenta region at $\nu = 1$ with $(S_3, C_3) = (0, +3)$, and the second is 
the cyan region at $\nu = 2$, with $(S_6, C_6) = (+3, -3)$.  These regions are independent
of $k_T a$ because the color-orbit coupling $k_T$ can be gauged away for $h_x/t = 0$,
that is, the Hamiltonian exhibits a color-gauge symmetry.
This leads to an energy spectrum that is independent of $k_T a$
(see for example Figs.~\ref{fig:one}a and~\ref{fig:one}b), and therefore the phase diagrams 
for all $k_T a$ are identical to the phase diagram for $k_T a = 0$.
At $h_x/t = 0$, the color-gauge symmetry of the Hamiltonian leads to 
topological numbers $S_r$ and $C_r$ that are independent of the color-orbit coupling
parameter $k_T a$.

In Fig.~\ref{fig:three}b, where $\alpha = 1/3$ and $h_x/t = 1$, there are many more topological
insulating phases and additional filling factors. 
At $\nu = 1/3$, there are red regions with $(S_1, C_1) = (0, +1)$ and
green regions with $(S_1, C_1) = (1, -2)$.
At $\nu = 2/3$, there are blue regions with $(S_2, C_2) = (0, +2)$.
At $\nu = 1$, there are very thin dark magenta regions $(S_3, C_3) = (-1, +6)$
close to $k_T a = 0.17\pi$ and $k_T a = 1.83\pi$ that have direct topological quantum
phase transitions to the central magenta region with $(S_3, C_3) = (0, +3)$.
At $\nu = 4/3$, there are small red-with-black-dots regions with $(S_4, C_4) = (+1, +1)$.
At $\nu = 5/3$, there are small yellow-with-black-dots regions with $(S_5, C_5) = (+2, -1)$.
At $\nu = 2$, there are very thin dark cyan regions $(S_6, C_6) = (+4, -6)$
close to $k_T a = 0.17\pi$ and $k_T a = 1.83\pi$ that have direct topological quantum
phase transitions to the central cyan region with $(S_6, C_6) = (+3, -3)$.
At $\nu = 7/3$, there are green regions with $(S_7, C_7) = (+3, -2)$.
At $\nu = 8/3$, there are yellow regions with $(S_8, C_8) = (+3, -1)$ and blue regions with 
$(S_8, C_8) = (+2, +2)$.

In Fig.~\ref{fig:three}c, where $\alpha = 1/3$ and  $h_x/t = 2$, the phase diagram is even richer than
in the case of Fig.~\ref{fig:three}b.
At $\nu = 1/3$, there are red regions with $(S_1, C_1) = (0, +1)$ and
green regions with $(S_1, C_1) = (1, -2)$,
and direct topological quantum phase transitions between them
at $k_T a =0.81\pi$ and $k_T a = 1.19\pi$.
At $\nu = 2/3$, there are blue regions with $(S_2, C_2) = (0, +2)$.
At $\nu = 1$, there are orange regions with  $(S_3, C_3) = (+1, 0)$ and magenta regions with
$(S_3, C_3) = (0, +3)$. The orange regions are very special, because they have two chiral midgap edge
states with opposite chirality, and although they do not produce a quantum charge Hall (QChH) effect,
they possess a quantum color Hall (QCoH) effect, similar to the quantum spin Hall effect
for spin-$1/2$ fermions.
At $\nu = 4/3$, there are red-with-black-dots and dark red regions with
$(S_4, C_4) = (+1, +1)$ and dark green and green regions with 
$(S_4, C_4) = (+2, -2)$.
There are direct topological quantum phase transitions between the red-with-black-dots and dark green regions
at $k_T a =0.25\pi$ and $k_T a = 1.75\pi$. The red-with-black-dots regions have three chiral midgap
edge states, two with positive and one with negative chirality, thus this phase possesses simultaneously
quantum charge Hall (QChH) and quantum color Hall (QCoH) effects, a situation that has no
correspondence for spin-$1/2$ fermions.
The dark green regions have two chiral midgap edge states with negative chirality
and achiral midgap edge states. When the achiral midgap edge states merge into the bulk, 
the dark green regions cross over to the green regions.
There are also direct topological quantum phase transitions between the dark green and dark red regions
at $k_T a =0.90\pi$ and $k_T a = 1.10\pi$. The dark red region has one chiral midgap edge state
with positive chirality, and at least one achiral midgap edge state.

The rest of the phases of Fig.~\ref{fig:three}c can be analyzed using particle-hole symmetry.
At $\nu = 5/3$, there are yellow-with-black-dots and dark yellow regions with $(S_5, C_5) = (+2, -1)$
and light blue and blue regions with  $(S_5, C_5) = (+1, +2)$. 
There are direct topological quantum phase transitions between the yellow-with-black-dots and light blue regions
at $k_T a =0.25\pi$ and $k_T a = 1.75\pi$. The yellow-with-black-dots regions have three chiral midgap
edge states, two with negative and one with positive chirality. thus this phase possesses simultaneously
quantum charge Hall (QChH) and quantum color Hall (QCoH) effects, a situation that has
no correspondence for spin-$1/2$ fermions.
The light blue regions have two chiral midgap egde states with positive chirality
and achiral midgap edge states. When the
achiral midgap edge states merge into the bulk, the light blue regions cross over to the blue regions.
There are also direct topological quantum phasse transitions between the light blue and dark yellow regions at
at $k_T a =0.90\pi$ and $k_T a = 1.10\pi$. The dark yellow regions have one chiral midgap edge state
with negative chirality, and at least one achiral midgap edge state.
At $\nu = 2$, there are orange regions with $(S_6, C_6) = (+2, 0)$ and
cyan regions with $(S_6, C_6) = (+3, -3)$.
Again, the orange regions are very special, because they have two chiral midgap edge
states with opposite chirality, and although they do not produce a quantum charge Hall (QChH) effect,
they possess a quantum color Hall (QCoH) effect, similar to the quantum spin Hall effect
for spin-$1/2$ fermions.
At $\nu = 7/3$, there are green regions with $(S_7, C_7) = (+3, -2)$.
At $\nu = 8/3$, there are yellow regions with $(S_8, C_9) = (+3, -1)$ and
blue regions with $(S_8, C_8) = (+2, +2)$, and
direct topological quantum phase transitions between them
at $k_T a =0.81\pi$ and $k_T a = 1.19\pi$.

In Fig.~\ref{fig:three}d, where $\alpha = 1/3$ and $h_x/t = 3$,
the phase diagram is as rich as in  Fig.~\ref{fig:three}c.
The main differences between the two figures
are the emergence of very small light brown regions at $\nu = 2/3$ with
$(S_2, C_2) = (-1,+5)$ near $k_T a = \pi$;
the emergence of cyan regions with $(S_3, C_3) = (+2, -3)$
and black regions with $(S_3, C_3) = (+1, 0)$ at $\nu = 1$;
the disappearance of the red-with-black-dot regions
$(S_4, C_4) = ( +1, +1)$ and
the merger of green regions with  $(S_4, C_4) = (+2, -2)$ at $\nu = 4/3$.
Similar effects occur to the phases with $\mu > 0$ by particle-hole symmetry
around $\mu = 0$, that is, for the mapping $\nu = r/3\to 3 - \nu$ with $C_r \to - C_{9 - r}$.
At $\nu = 1/3$, there are
red regions with $(S_1, C_1) = (0, +1)$ and
green regions with $(S_1, C_1) = (+1, -2)$,
and direct topological quantum phase transitions between them
at $k_T a = 0.82\pi$ and $k_T a = 1.18\pi$.
At $\nu = 2/3$, there are blue regions with $(S_2, C_2) = (0, +2)$ and
very small light brown regions $(S_2, C_2) = (-1,+5)$ near $k_T a = \pi$
that have a direct topological quantum phase transition into also very small new blue regions
at $k_T a = 0.90\pi$ and $k_T a = 1.10\pi$.
At $\nu = 1$, there are
orange regions with  $(S_3, C_3) = (+1, 0)$,
cyan regions with $(S_3, C_3) = (+2, -3)$
and black regions with $(S_3, C_3) = (+1, 0)$.
There are also direct topological quantum phase transitions from the
orange to the cyan regions at $k_T a = 0.31\pi$ and $k_T a = 1.69\pi$, as well as,
from cyan to black regions at at $k_T a = 0.46\pi$ and $k_T a = 1.54\pi$. 
At $\nu = 4/3$, there are
green regions with $(S_4, C_4) = (+2, -2)$,
as well as red and dark red regions with $(S_4, C_4) = (+1, +1)$.
There are direct topological quantum phase transitions between
the green and red regions at $k_T a = 0.74\pi$ and $k_T a = 1.24\pi$.
Notice that the dark red region crosses over into the red region (around $k_T a = \pi$),
when all its achiral midgap edge states merge into the bulk.

The additional phases of Fig.~\ref{fig:three}d reflect particle-hole symmetry about $\mu = 0$.
At $\nu = 5/3$, there are blue regions with $(S_5, C_5) = (+1, +2)$
and yellow and dark yellow regions with  $(S_5, C_5) = (+2, -1)$.
There are direct topological quantum phase transitions between the
blue and yellow regions at $k_T a = 0.74\pi$ and $k_T a = 1.26\pi$.
Notice that the dark yellow region crosses over into the yellow region (around $k_T a = \pi$),
when all its achiral midgap edge states merge into the bulk.
At $\nu = 2$, there are orange regions with $(S_6, C_6) = (+2, 0)$,
magenta regions with $(S_6, C_6) = (+1 +3)$
and black regions with $(S_6, C_6) = (+2, 0)$.
There are also direct topological quantum phase transitions from the
orange to the magenta regions at $k_T a = 0.31\pi$ and $k_T a = 1.69\pi$, as well as,
from the magenta to black regions at at $k_T a = 0.46\pi$ and $k_T a = 1.54\pi$. 
At $\nu = 7/3$, there are green regions with $(S_7, C_7) = (+3, -2)$ and
very small light pink regions $(S_7, C_7) = (+4,-5)$ near $k_T a = \pi$ 
that have direct topological quantum phase transitions into very small new green regions
at $k_T a = 0.90\pi$ and $k_T a = 1.10\pi$.
At $\nu = 8/3$, there are yellow regions with $(S_8, C_8) = (+3, -1)$ and blue regions
with $(S_8, C_8) = (+2, +2)$, and direct topological quantum phase transitions between them
at $k_T a = 0.82\pi$ and $k_T a = 1.18\pi$.

From the analysis of the phase diagrams in Figs.~\ref{fig:three} and~\ref{fig:four}, we have established
that there is a staircase of filling factor $\nu$ versus chemical potential $\mu$ labeled by topological
quantum numbers $(S_r, C_r)$ at every gap $r$ in the energy spectrum.
The steps of the staircase structure 
occur at values of $\nu$ given by the gap labelling theorem displayed in Eq.~(\ref{eqn:gap-labelling}).
To make this connection more evident, we analyze next the color density
of states $\rho_c (E)$ and the total color density of states $\rho (E) = \sum_c \rho_c (E)$.
From $\rho (E)$, we compute directly the filling factor $\nu$ as a function of chemical potential $\mu$
and show that there are steps in the function $\nu (\mu)$
at the precise values determined by the gap labelling theorem.
This indicates the existence of incompressible insulating phases labeled
by the topological indices $(S_r, C_r)$ at $\nu = r/q$, as expected.

\section{Color Density of States}
\label{sec:color-density-of-states}

In conjunction with the energy spectrum $E_{n_\beta} (k_y)$ with open boundary
conditions or $E (k_x, k_y)$ with periodic boundary conditions for fixed magnetic flux $\alpha = p/q$,
the total color density of states $\rho (E)$ and the color density of states $\rho_c (E)$ for color $c$
are useful quantities to identify the location of gapped phases as a function of color-flip
fields $h_x/t$ and color-orbit coupling $k_T a.$

The density of states can be obtained from the Green  (Resolvent) operator
\begin{equation}
{\hat {\bf G}} (z)  =
\frac{1}{z {\bf 1} - {\hat {\bf H}}},
\end{equation}
whose matrix elements in the original color basis $\{ R, G, B \}$ can be written as
\begin{equation}
G_{c c^\prime} (z)
=
\sum_{n_{\beta} k_y}
\frac{u_{n_{\beta} c} (k_y) u_{n_\beta  c^{\prime}}^* (k_y)}{z - E_{n_{\beta}} (k_y)}  
\end{equation}
where $u_{n_\beta  c} (k_y)$ are the color components of the eigenvectors
of the Hamiltonian operator ${\hat {\bf H}}$ with open boundary conditions
and eigenvalues $E_{n_\beta} (k_y)$.
The summations over $k_y$ cover the magnetic Brillouin zone $[-\pi/qa, \pi/qa]$ for
bulk states and the range $[-\pi/a, \pi/a]$ for midgap edge states, where $a$ is
the square lattice unit cell length, and
include all mixed-color band indices $n_{\beta}.$
Using the appropriate spectral decomposition for the case with periodic
boundary conditions, we obtain
\begin{equation}
G_{c c^\prime} (z)
=
\sum_{\ell_\gamma k_x  k_y}
\frac{u_{\ell_\gamma c} (k_x , k_y) u_{\ell_\gamma  c^{\prime}}^* (k_x ,k_y)}{z - E_{\ell_\gamma} (k_x . k_y)} ,
\end{equation}
where $u_{\ell_\gamma c} (k_x , k_y)$ are the color components of the eigenvectors
of the Hamiltonian operator ${\hat {\bf H}}$ with periodic boundary conditions
and eigenvalues $E_{\ell_{\gamma}} (k_x, k_y)$.
Here, $\ell_{\gamma}$ labels the magnetic subbands for $\alpha = p/q$.
The momentum summations are over $[-\pi/a, \pi/a]$ for $k_x$ and
over $[-\pi/qa, \pi/qa]$ for $k_y$, that is,
the summations over $\{\ell_{\gamma}, k_x, k_y\}$ cover the magnetic Brillouin zone
and all the mixed color bands labelled by $\ell_\gamma$.

Within the magnetic Brillouin zone, 
the density of states of color $c$ at energy $E$ is 
\begin{equation}
{\bar \rho}_{c} (E) =
 -\frac{1}{\pi} \lim_{\delta \to 0} {\rm Im} G_{c c} (z = E + i\delta),
\end{equation}
where $\delta$ is a small imaginary part. The color density of states per site is 
\begin{equation}
 {\rho}_{c} (E) =
 \frac{{\bar \rho}_c (E)}{q},
\end{equation}
since there are $q$ unit cells in real space.
The color density of states ${\bar \rho}_{c} (E)$ in the magnetic unit cell
integrates to $q$ states over all energies. The color density of states per site
${\rho}_{c} (E)$ always integrates to $1$, because
we have a maximum of one state for a given color $c$. 
Within the magnetic Brillouin zone, the number of states of a given color $c$
at the chemical potential $\mu$ is 
\begin{equation}
N_c (\mu)  = \int_{\rm E_{min}}^{\mu}dE {\bar \rho}_c (E),
\end{equation}
where $E_{\rm min}$ is the minimum energy in the spectrum.
The maximum value of $N_c (\mu)$ is $N_{c, {\rm max}} = q$, since there is maximum
of one color state $c$ per site, and $q$ is the number of sites contained in the real space magnetic
unit cell. The filling factor for color $c$ is defined as the ratio
\begin{equation}
\nu_c (\mu) = \frac{N_c (\mu)}{N_{c, {\rm max}}},
\end{equation}
which has a maximal value of one, that is, $\nu_{c, {\rm max}} = 1$.

The total density of states within the magnetic unit cell can be written as 
\begin{equation}
{\bar \rho} (E) = \sum_c {\bar \rho}_{c} (E),
\end{equation}
while the total density of states per site has the form
\begin{equation}
 {\rho} (E) =
 \frac{{\bar \rho}(E)}{q}.
\end{equation}
The total density of states ${\bar \rho} (E)$, within the unit cell, integrates to $3q$ states
over all energies, while the density of states per site ${\rho} (E)$ always integrates to $3$,
because we have a maximum of three colors per site. The total filling factor
at chemical potential $\mu$ is 
\begin{equation}
\nu (\mu) = \sum_c \nu_c (\mu),
\end{equation}
having a maximum value $ \nu_{\rm max}= 3$.

\begin{figure} [tb]
\centering 
\epsfig{file=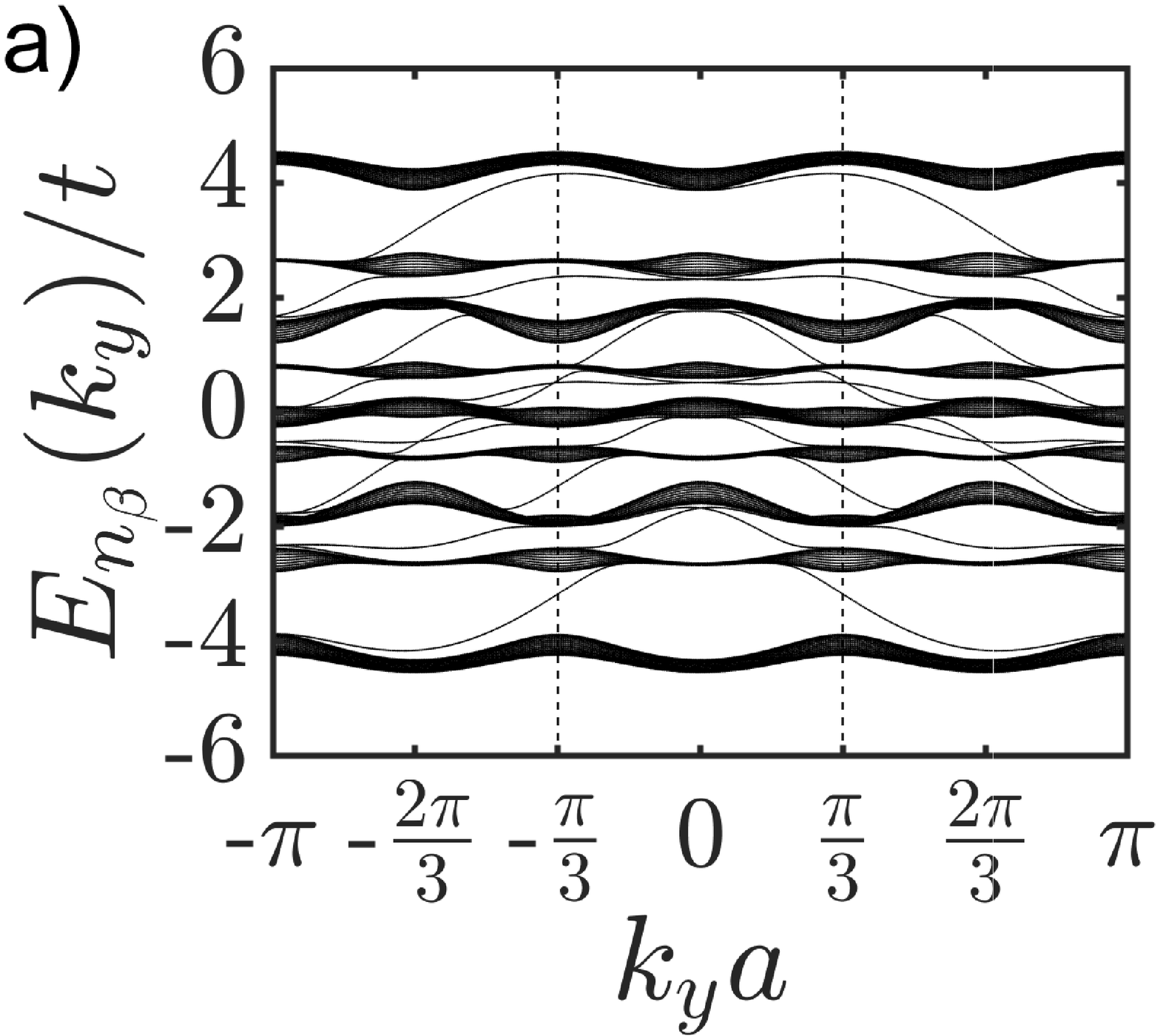,width=0.49 \linewidth}
\epsfig{file=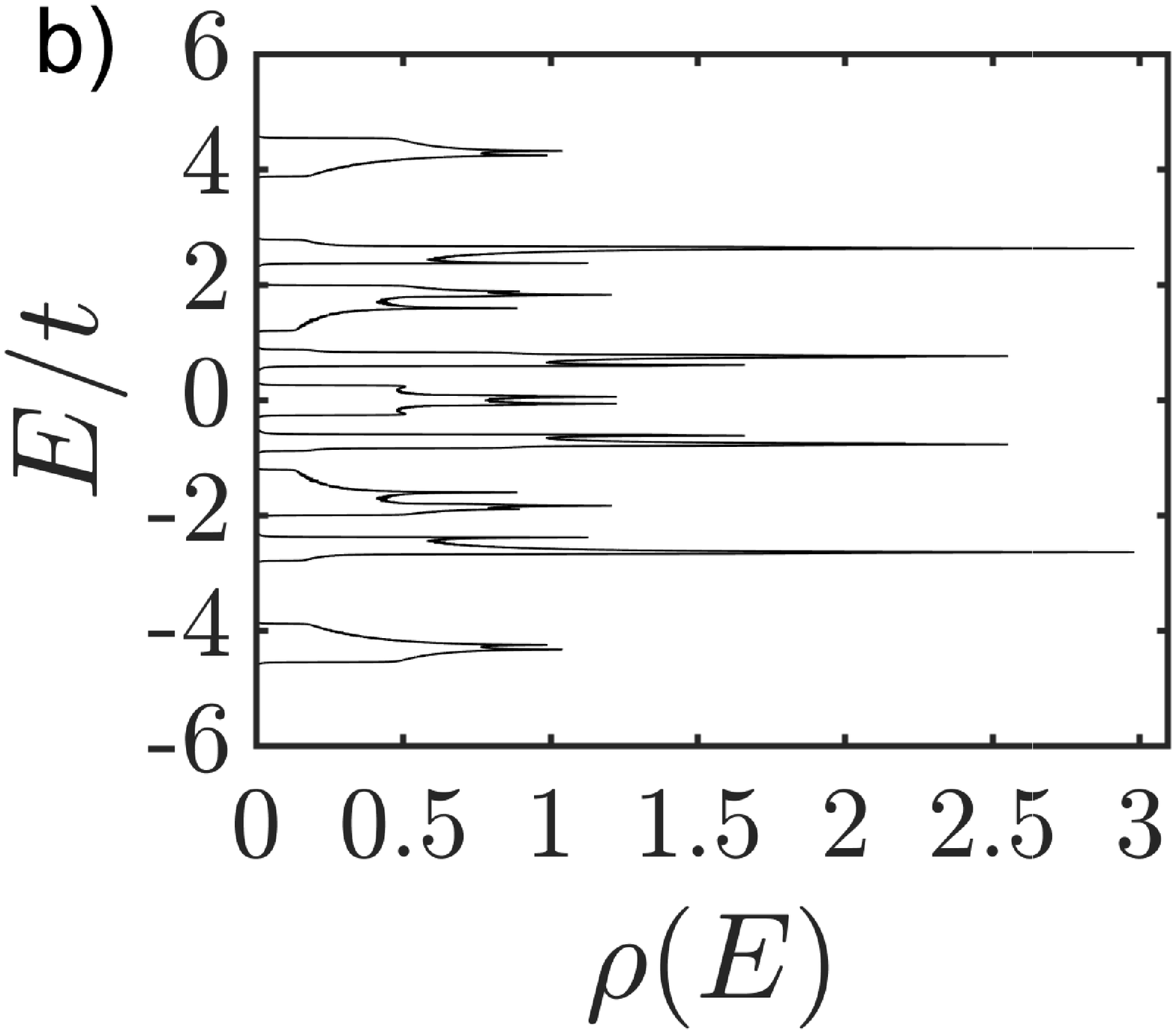,width=0.49 \linewidth}
\vskip 0.2cm
\epsfig{file=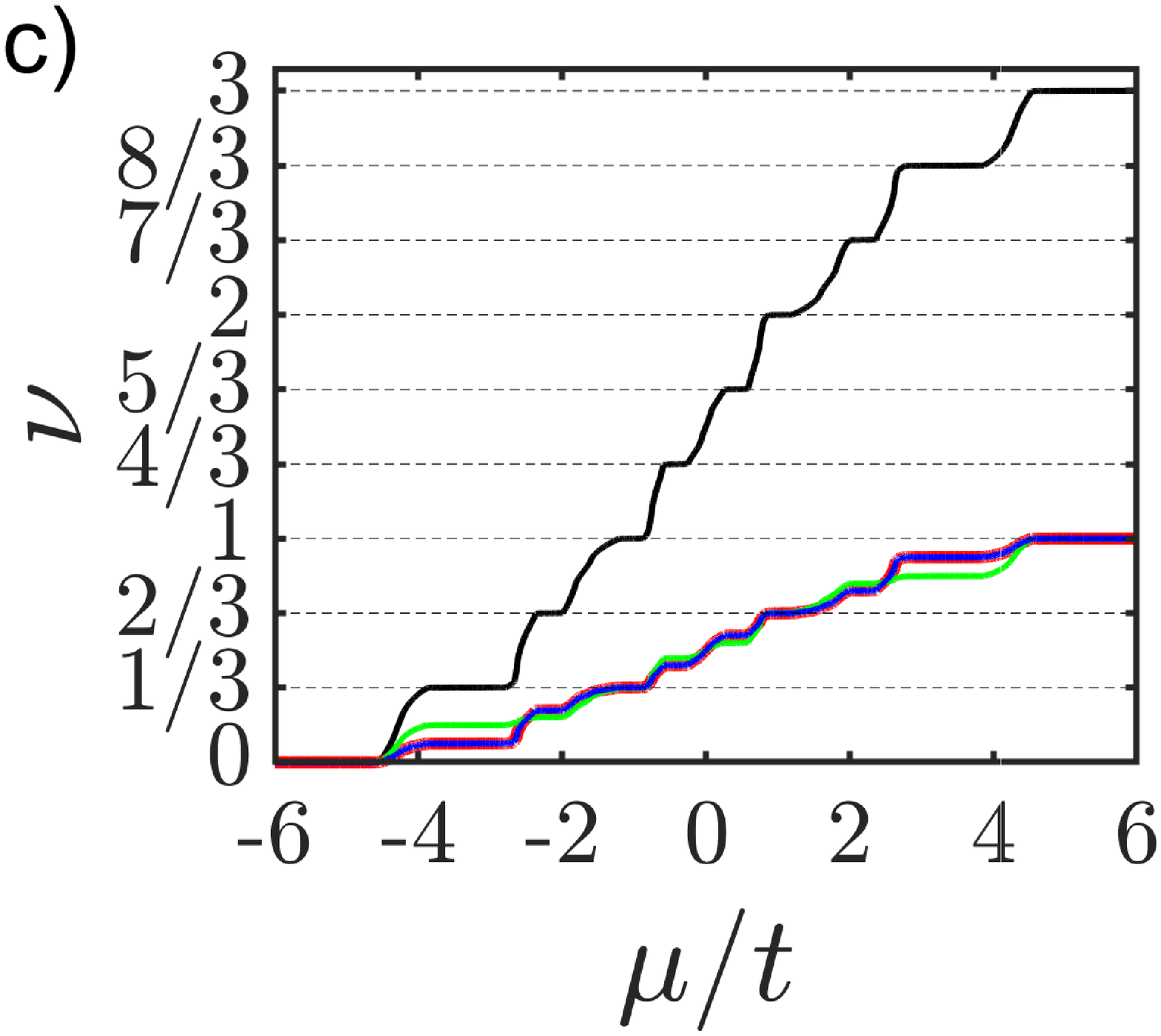,width=0.49 \linewidth}
\epsfig{file=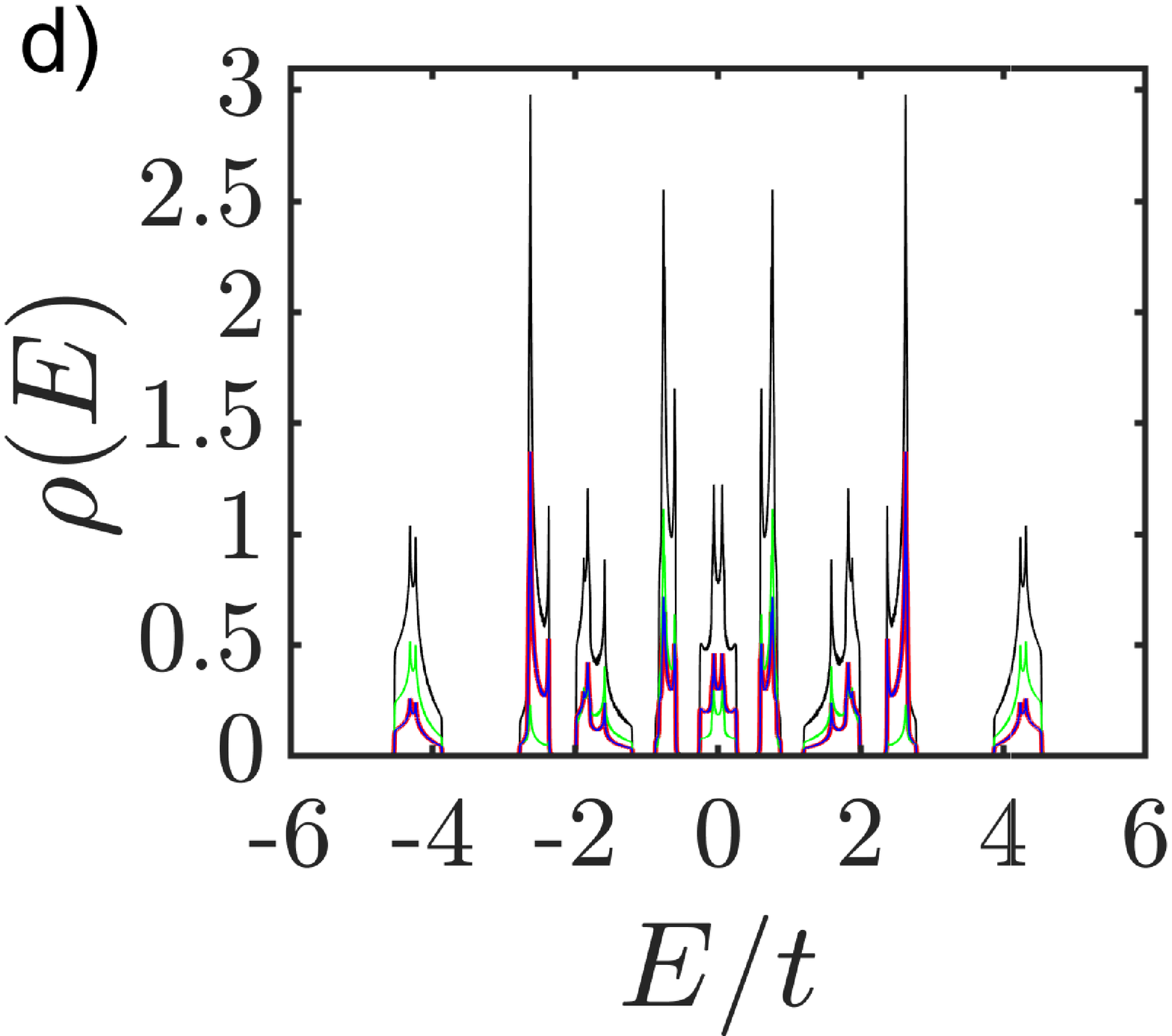,width=0.49 \linewidth}
\caption{ 
\label{fig:four}
(Color Online)
Spectroscopic properties and filling factor for parameters
$\alpha = 1/3$, $k_T a = \pi/8$ and $h_x/t = 1.85$.
This corresponds to a vertical line across the phase diagram of Fig.~\ref{fig:two}b.
a) Energy spectrum $E_{n_\beta} (k_y )/t$ versus $k_y a$ for the case of open boundary conditions,
showing explicitly midgap edge states.
The panels b), c) and d) refer to the case of periodic boundary conditions.
b) Energy $E/t$ versus total density of states per site $\rho (E)$ to illustrate the gaps
between the bands, which coincide with the gaps for the case of open boundary conditions.
c) Filling factor $\nu$ versus chemical potential $\mu/t$ showing steps where incompressible
insulating phases occur.
d) Density of states per site $\rho (E)$ versus energy $E/t$. The total color density states
per site $\rho (E)$
is in black, and the color density of states are in red for the Red ${\rm (R)}$  states,
in green for the Green ${\rm (G)}$ states and in blue for the Blue ${\rm (B)}$ states. 
}
\end{figure}

In Fig.~\ref{fig:four}, we show spectroscopic information for particular values of parameters:
the magnetic flux ratio $\alpha = 1/3$, the color-flip field $h_x/t = 1.85$ and the color-orbit coupling
parameter $k_T a = \pi/8$. All these parameters are the same for
Figs.~\ref{fig:four}a through~\ref{fig:four}d. The choice of these parameters illustrates a 
vertical scan at $h_x/t = 1.85$ in Fig.~\ref{fig:two}b. The vertical line cuts through a wide variety
of topological phases as the chemical potential $\mu$ grows.

In Fig.~\ref{fig:four}a, we show the energy eigenspectrum versus momentum $k_y a$
for open boundary conditions, where eight gaps can be seen in the spectrum corresponding to
the insulating phases of the vertical scan in Fig.~\ref{fig:two}b at $h_x/t = 1.85$.
The energy dispersions of the midgap edge states are also shown.
The first gap $(r = 1)$ corresponds to the red region in Fig.~\ref{fig:two}b with
one chiral midgap edge state with positive chirality
and topological numbers $(S_1, C_1) = (0, +1)$.
The second gap $(r = 2)$ corresponds to the blue region in Fig.~\ref{fig:two}b with
two chiral midgap edge states with positive chirality
and topological numbers $(S_2, C_2) = (0 , +2)$.
The third gap $(r = 3)$ correponds to the orange region in Fig.~\ref{fig:two}b with
two chiral midgap edge states with opposite chirality
and topological numbers $(S_3, C_3) = (+1 , 0)$.
The fourth gap $(r = 4)$ correponds to the red-with-black-dots region in Fig.~\ref{fig:two}b with
two chiral midgap edge states with positive chirality,
one chiral midgap state with negative chirality
and topological numbers $(S_4, C_4) = (+1 , +1)$.
The fifth  gap $(r = 5)$ correponds to the yellow-with-black-dots region in Fig.~\ref{fig:two}b with
two chiral midgap edge states with negative chirality,
one chiral midgap state with positive chirality
and topological numbers $(S_5, C_5) = (+2 , -1)$.
The sixth gap $(r = 6)$ correponds to the orange region in Fig.~\ref{fig:two}b with
two chiral midgap edge states with opposite chirality
and topological numbers $(S_6, C_6) = (+2 , 0)$.
The seventh gap $(r = 7)$ corresponds to the green region in Fig.~\ref{fig:two}b with
two chiral midgap edge states with negative chirality
and topological numbers $(S_7, C_7) = (+3 , -2)$.
The eighth  gap $(r = 8)$ corresponds to the yellow region in Fig.~\ref{fig:two}b with
one chiral midgap edge state with negative chirality
and topological numbers $(S_8, C_8) = (+3, -1)$.

In Fig.~\ref{fig:four}b, we show a plot of energy $E$ versus total color density of states
per site $\rho (E)$ for periodic boundary conditions to indicate explicitly the location
of the gaps in the energy spectrum. We used a small imaginary part $(\delta = 5 \times 10^{-3} t)$
to calculate $\rho (E)$ from $G_{cc} (z = E + i\delta)$.
The eight energy gaps can be clearly seen at the locations where the total color density of
states in the bulk is zero. The regions associated with these gaps correspond to the
eight phases that are crossed in a vertical scan in Fig.~\ref{fig:two}b at $h_x/t = 1.85$
as the chemical potential $\mu$ or filling factor $\nu$ grows.

In Fig.~\ref{fig:four}c, we show a plot of the color filling factors $\nu_c $ and
the total filling factor $\nu$ versus chemical potential $\mu$,
calculated using the color density of states per site $\rho_c (E)$ and the
total color density of states per site $\rho (E)$, respectively.  Notice that when
the chemical potential $\mu$ lies inside a band gap, the filling factor $\nu$
is constant and take the exact form $\nu = r/q$ as discussed in connection
to the gap labelling theorem of Sec.~\ref{sec:gap-labelling-theorem}.
The gapped phases (the chemical potential $\mu$ lies inside of a gap) are incompressible,
because the total color filling factor $\nu$ and the color filling factor $\nu_c$ for each color
is constant, and the color compressibilty $\kappa_c$ is proportional to $d\nu_c (\mu)/ d\mu = 0$.
Naturally, the sequence of insulating phases crossed as $\mu$ increases is exactly the same as
that in Fig.~\ref{fig:two}b.
At $\nu = 0$ and $\nu = 3$, we have trivial insulating phases. 
At $\nu = 1/3$, there is a red region with topological quantum numbers $(S_1, C_1)  = (0, +1).$   
At $\nu = 2/3$, there is a blue region with topological quantum numbers $(S_2, C_2) = (0, +2)$.
At $\nu = 1$, there is an orange region with topological quantum numbers $(S_3, C_3)  = (+1, 0)$.
At $\nu = 4/3$, there is a red-with-black-dots region with topological quantum numbers
$(S_4, C_4) = (+1, +1).$   
At $\nu = 5/3$, there is a yellow-with-black-dots region with topological quantum numbers
$(S_5, C_5)  = (+2, -1).$
At $\nu = 2$, there is an orange region with topological quantum numbers $(S_6, C_6) = (+2, 0)$.
At $\nu = 7/3$, there is a green region with topological quantum numbers $(S_7, C_7) = (+3, -2)$.
At $\nu = 8/3$, there is a yellow region with topological quantum numbers $(S_1, C_1) = (+3, -1).$
Notice also that filling factors $\nu_R = \nu_B$ reflecting a symmetry of the Hamiltonian operator
in Eq.~(\ref{eqn:hamiltonian-matrix}) via the simultaneous exchange $R \leftrightarrow B$ and
$k_T \leftrightarrow - k_T$.

In Fig.~\ref{fig:four}d, we show a plot of the color density of states per site $\rho_c$  and the
total color density of states per site $\rho$ versus energy $E$.
Notice that $\rho_R (E) = \rho_B (E)$, reflecting a symmetry of the Hamiltonian operator
in Eq.~(\ref{eqn:hamiltonian-matrix}) via the simultaneous
exchange $R \leftrightarrow B$ and $k_T \leftrightarrow - k_T$.
One can clearly see the eight bands characterizing the insulating states
discussed in Fig.~\ref{fig:four}c.
The total color density states per site $\rho (E)$
is in black, and the color density of states are in red for the Red ${\rm (R)}$  states,
in green for the Green ${\rm (G)}$ states and in blue for the Blue ${\rm (B)}$ states. 

\section{Summary and Conclusions}
\label{sec:summary-and-conclusions}

We investigated the eigenspectrum, Chern numbers and phase diagrams
of ultracold color-orbit coupled ${\rm SU(3)}$ fermions in optical lattices, having
in mind possible experimental systems, such as fermionic isotopes
$^{137}{\rm Yb}$ and $^{87}{\rm Sr}$. We labeled the internal states of the atoms by
colors Red (R), Green (G) and Blue (B), and analyzed the quantum phases as a function
of artificial magnetic, color-orbit and color-flip fields that can be independently controlled.

For fixed artificial magnetic flux ratio, we identified topological quantum phases and
phase transitions in the phase diagrams of chemical potential versus color-flip fields or
color-orbit coupling, where the chirality and number of
midgap edge states change. We established a gap labelling theorem to characterize
the insulating phases by their filling factors and topological quantum numbers.

The topologically non-trivial phases are classified in three groups:
the first group has total non-zero chirality and exhibit only the quantum charge Hall effect;
the second group has total non-zero chirality and exhibit both quantum charge and quantum
color Hall effects;
and the third group has total zero chirality, but exhibit the quantum color Hall effect.
These phases are generalizations of the
quantum Hall and quantum spin Hall phases for charged spin-$1/2$ fermions.

Lastly, we described the color density of states per site and a staircase structure in the total
and color filling factors versus chemical potential for fixed color-orbit,
color-flip and magnetic flux ratio. We showed the existence of incompressible states 
at rational filling factors precisely given by a gap labelling theorem that related the filling
factors to the magnetic flux ratio and topological quantum numbers.

Our theoretical findings pave the way for the experimental discovery of topological insulating
phases that present simultaneously a quantum charge Hall effect (QChH)  and a quantum color Hall
effect (QCoH) in ${\rm SU(3)}$ fermions such as $^{173}{\rm Yb}$ or $^{87}{\rm Sr}$.
This particular phase has no correspondence for spin-1/2 fermions in condensed matter
or ultracold atomic physics, where the quantum Hall and the quantum spin Hall phases
are mutually exclusive.

\acknowledgements{
One of us (C.A.R.S.d.M.) would like to thank the support of 
the Galileo Galilei Institute for Theoretical Physics (Florence, Italy)
via a Simons Fellowship, and of the International Institute of Physics (Natal, Brazil)
via its Visitor's Program. 
}

\end{document}